\documentclass{aa}  
\usepackage{comment}

\usepackage[utf8]{inputenc}
\usepackage[dvipsnames]{xcolor}
\usepackage{soul} 
\usepackage{adjustbox}
 
\usepackage{txfonts}

\usepackage{placeins}
\usepackage{float}
\usepackage{graphicx}	
\usepackage{amsmath}	
\usepackage{amssymb}	
\usepackage{algorithm} 
\usepackage{algpseudocode} 
\usepackage{bm}
\usepackage[colorlinks=true,
    linkcolor=blue,
    citecolor=blue,
    filecolor=magenta,      
    urlcolor=cyan]{hyperref}

\newcommand{\lsim}{\mathrel{\rlap{\lower 3pt \hbox{$\sim$}} \raise 2.0pt \hbox{$<$}}}
\newcommand{\gsim}{\mathrel{\rlap{\lower 3pt \hbox{$\sim$}} \raise 2.0pt \hbox{$>$}}}

\newcommand{\fr}[1]{\textcolor{black}{ #1}}

\begin{document} 

   \title{ESPRESSO reveals a single but perturbed broad-line region in the supermassive black hole binary candidate PG~1302-102} 
   \titlerunning{The perturbed BLR of PG~1302-102   }
   \authorrunning{F. Rigamonti et al.}

   \author{Fabio Rigamonti
          \inst{1,2,}
          \inst{12} \fnmsep\thanks{fabio.rigamonti@inaf.it},
          Paola Severgnini\inst{1},
          Erika Sottocorno\inst{3},
          Massimo Dotti\inst{3, 1, 2},
          Stefano Covino\inst{1,12},
          Marco Landoni\inst{1},
          Lorenzo Bertassi\inst{3},
          Valentina Braito\inst{1,7,8},
          Claudia Cicone\inst{4},
          Guido Cupani\inst{5,6},
          Alessandra De Rosa\inst{11},
          Roberto Della Ceca\inst{1},
          Luca Ighina\inst{1},
          Jasbir Singh\inst{1},
          \and
          Cristian Vignali\inst{9,10}
    }

   \institute{
            INAF - Osservatorio Astronomico di Brera, via Brera 20, I-20121 Milano, Italy
        \and
            INFN, Sezione di Milano-Bicocca, Piazza della Scienza 3, I-20126 Milano, Italy
        \and
            Universit\`a degli Studi di Milano-Bicocca, Piazza della Scienza 3, 20126 Milano, Italy
        \and
            Institute of Theoretical Astrophysics, University of Oslo, PO Box 1029, Blindern 0315, Oslo, Norway
        \and 
            INAF -- Osservatorio Astronomico di Trieste, via Tiepolo 11, I-34143 Trieste, Italy
        \and 
            Institute for Fundamental Physics of the Universe, via Beirut 2, I-34151 Trieste, Italy
        \and
            Dipartimento di Fisica, Università di Trento, Via Sommarive 14, Trento 38123, Italy
        \and
            Department of Physics, Institute for Astrophysics and Computational Sciences, The Catholic University of America, Washington, DC 20064, USA
        \and
        Dipartimento di Fisica e Astronomia 'Augusto Righi', Universit\`a degli Studi di Bologna, Via Gobetti 93/2, I-40129 Bologna, Italy 
        \and
        INAF -- Osservatorio di Astrofisica e Scienza dello Spazio di Bologna, Via Gobetti 93/3, I-40129 Bologna, Italy
        \and
        INAF - Istituto di Astrofisica e Planetologia Spaziali (IAPS), via Fosso del Cavaliere, Roma, I-133, Italy
        \and
        Como Lake centre for AstroPhysics (CLAP), DiSAT, Università dell’Insubria, via Valleggio 11, 22100 Como, Italy
    }

   \date{Received XXX; accepted YYY}

\abstract{
In this work, we present a new, fully Bayesian analysis of the highest-resolution optical spectrum of the supermassive black hole (SMBH) binary candidate PG~1302-102, obtained with ESPRESSO at the VLT (\fr{$\rm{R}\simeq 138,000$}). Our methodology, based on robust Bayesian model selection, reveals the presence of multiple narrow emission lines at the expected redshift of the source and confirms (for H$\beta$) and detects (for H$\gamma$) the clear presence of redshifted broad components. Additionally, we have discovered a ``very broad'' and, if it is associated with the H$\beta$, ``very redshifted'' component at $\lambda\simeq5000\AA$. We evaluate two scenarios for explaining the observed broad emission line (BEL) features in PG~1302-102. In the case in which the redshifted BEL asymmetry arises from the orbital motion of a putative binary, our measurements coupled with simple estimates of the broad-line region (BLR) sizes suggest that the individual black hole BLRs are either settled in a single BLR or in the process of merging and, therefore truncated and highly disturbed. Alternatively, in the scenario of a single SMBH, we explain the distorted emission of the BELs with a nonsymmetric distribution of the BLR clouds; namely, a thin disk with a spiral perturbation. This BLR configuration is statistically preferred over any empirical multi-Gaussian fit and simultaneously explains the asymmetric emission of the H$\beta$ and H$\gamma$ close to the bulk of
the line and any additional excess (or the lack of it, in the case of the H$\gamma$) at much longer wavelengths. The physical origins of the perturbation are still unclear and a connection with the possible presence of a black hole binary cannot be ruled out. Given the growing evidence from theoretical and observational works demonstrating the common presence of disturbed BLRs in active galactic nuclei, we argue that an origin related to self-gravitating instabilities may be more plausible.}

\keywords{galaxies: active - galaxies: interactions -  quasars: individual: PG~1302-102 - quasars: emission lines - quasars: supermassive black holes - Techniques: spectroscopic}

\maketitle

\section{Introduction}
\label{sec:intro}
Supermassive black hole (SMBH) binaries (SMBHBs) are thought to be common in the Universe as a natural outcome of galaxy mergers \citep[][]{BBR80}. Such systems are among the loudest sources of gravitational waves (GWs) potentially detectable with current pulsar timing array campaigns \citep{pta,Agazie23} and future space-based GW interferometers \citep[e.g., LISA,][]{lisa1,lisa2}. However, theoretical expectations on the rates of detectable binary mergers are still highly uncertain; such uncertainties could in principle be reduced through the electromagnetic identification of bona fide SMBHBs \citep{Derosa2019}.

Unfortunately, several challenges complicate the unambiguous identification of these sources and, to date, no definite observational confirmation of any SMBHB has been made yet. The only binary candidate, spatially resolved through very long baseline radio interferometry observations, is the radio galaxy 0402+379 \citep{Rodriguez09, BurkeSpolaor11} showing two flat-spectrum radio cores with a projected separation of $\approx 7$ pc (see also \citealt{Kharb2017} for a 0.35 pc binary in the spiral galaxy NGC 7674).  At smaller separations, unresolvable with current facilities except for the closest active galactic nuclei (AGNs), SMBHBs have been searched-for either through photometric variability in their light curve  (\citealt{Valtonen08, Ackermann15, Graham15, Li2016, Charisi16, Sandrinelli16,Sandrinelli18, Severgnini18, Li+2019, LiuGez+2019,Serafinelli+20, Chen+2020,Covinoetal2020}) or by looking for peculiar spectral features \citep{Tsalmantza11, Eracleous12, Ju13, Shen13, Wang17}.


If the two black holes are at a separation such that at least one of them still retains its own broad-line region (BLR), then we expect broad emission lines (BELs) to be shifted in frequency with respect to their respective narrow emission lines (NELs) and to evolve in time over a binary orbital period. Unfortunately, asymmetric emission line profiles could also be explained by eccentric BLRs \citep[][]{eracleous97}, as well as circular BLRs featuring deviations from an axisymmetric emissivity profile; for example, as hot spots \citep[][]{Jovanovic_2010} or spiral patterns  \citep[][]{storchi17}. In addition, even symmetric BLRs might appear as asymmetric in observations in the presence of partial obscuration from dust \citep[][]{Gaskell2018}, or if the BLR is attached to a recoiling SMBH \citep[][]{Volonteri2008}. In these cases, the obvious way of checking for the presence of a SMBHB would be to measure the expected drift in wavelengths by observing the candidate over a binary orbital period\footnote{Given the typical binary separations involved in order to be conclusive\fr{,} such test could require up to $\gsim 10 - 100$ yr \citep[][]{Tsalmantza11,Eracleous12}.}.  

At smaller separations where the BLR is either truncated \citep{Montuori11} or shared by both SMBHs, photometric variability is still expected over orbital timescales due to periodic fueling from the circumbinary material \citep[e.g.,][]{HMH08,Tiede2024}, Doppler boosted (DB) emission \citep{DHS15}, or periodic gravitational lensing from the companion of the active component of the SMBHB \citep[][]{doraziolense18, davelaar22a, davelaar22b}. Even in these cases, alternative interpretations have been proposed for quasi-periodically modulated sources, either assuming Lense-Thirring driven precessing jets \citep[][]{Sandrinelli16,Britzen2018} or only apparent (quasi-)periodicities due to the effect of correlated noise together with the intrinsic difficulties affecting the analysis of short (compared to the searched-for periodicities) time series \citep[][]{Vaughan2016,Liu2018Asasn,Covinoetal2019}.

One of the strongest candidate SMBHBs identified through periodicity in its light curve is PG~1302-102; a quasar with a median V-band magnitude of 15.0 and a redshift
of 0.2784 showing variability in multiple bands \citep{Liu2024} as well as the presence of BELs \citep[][]{Graham2015Nat}. This object has been observed across multiple wavelengths, including optical \citep[][]{Graham2015Nat}, ultraviolet \citep[][]{DHS15,xin20}, infrared \citep[][]{Jun2015}, and X-rays \citep[][]{Saade2024}\footnote{See also \citep[][]{Liu2024} for photometric reverberation mapping measurement obtained combining such light curves.}. The modulation of PG~1302-102 has been explained through the DB emission of an unequal-mass binary: when the secondary black hole accretes matter onto a mini-disk, the emission is DB due to the relativistic\footnote{Estimates from the periodicity and the predicted mass of the putative binary give a separation of $\simeq0.01\rm{pc}$ and, therefore relativistic motion.} orbital motion of the secondary. PG~1302-102 also exemplifies the well-known challenge of determining the intrinsic nature of a quasar based on the observation of a limited number of periodic cycles \citep[][]{Covinoetal2019}; indeed, its true nature remains a topic of debate. For instance, \cite{Vaughan2016} argue that the light curve observed in PG~1302-102 is better described with a damped random walk model \citep[][]{Kelly2009} rather than a sinusoid with uncorrelated noise \fr{\citep[see also][]{Graham2015Nat}}. This hypothesis is supported by the fact that PG~1302-102 was selected as the most periodic source from a sample of $\sim 250$ $000$ light curves. With such ratios (i.e., 1 in 250 000), it is likely that one will find some quasar that appears periodic purely by coincidence. Similarly, \cite{Liu2018Asasn} measured a decreased evidence for periodicity when additional data were included in the modeling. Conversely, \cite{Zhu2020} show that if correlated noise is combined with sinusoidal variation then quasi-periodic oscillation is the most favored explanation for PG~1302-102 variability.  

In this paper, we aim to further clarify the nature of PG 1302-102. We apply a fully Bayesian approach to the highest-resolution spectrum of this source, observed using the Echelle SPectrograph for Rocky Exoplanets and Stable Spectroscopic Observations \citep[ESPRESSO,][]{Pepe2021} on the Very Large Telescope (VLT). Specifically, we present an improved characterization of the broad and narrow line region properties of PG~1302-102, by examing the emission lines for substructures, such as nonsymmetric, redshifted, and blueshifted components.

This paper is organized as follows: in \S~\ref{sec:data_sample} we describe the collected data and the reduction process; in \S~\ref{sec:modelling} we detail our procedure for determining the best description of the spectrum also providing a physically motivated interpretation for the BLR of PG~1302-102; in \S~\ref{sec:discussion} we discuss the implication of our findings in relation with the putative binary nature of PG~1302-102; and, finally, in \S~\ref{sec:conlcusion} we provide a short summary of our work.

\section{Data}
\label{sec:data_sample}
PG~1302-102 was observed with VLT ESPRESSO in early 2018, during the instrument’s third commissioning run. Three exposures of 1200 s each were taken on February 27, February 28, and March 3. The instrument was operated in SINGLEHR mode, meaning that it was accepting light from VLT Unit Telescope (UT) at a time (respectively, UT2, UT1, and UT3 for the three exposures), using the 140 $\mu$m fibre for high resolution. The detector was binned by 2 in the spatial direction and by 1 in the spectral direction, achieving a resolving power of approximately \fr{$138,000$}.

The data were reduced with the official pipeline released by ESO\footnote{\url{https://www.eso.org/sci/software/pipelines/espresso/}}, version 1.3.2. Standard calibrations were used to remove the instrumental signature and calibrate the spectra into physical units. Wavelength calibration was performed using the instrument built-in Fabry-Pérot interferometer, with a ThAr lamp acting as an absolute reference. Flux calibration was performed with the aid of two spectro-photometric standard stars (HR 4963 for February 27 and March 3, HR 5501 for March 3). The final products of the pipelines were three reduced (flat-fielded, sky-subtracted, wavelength- and flux-calibrated) spectra, provided both in order-by-order and merged format.

To perform the analysis, we coadded the three reduced spectra using the ESPRESSO Data Analysis Software (DAS)\footnote{\url{https://www.eso.org/sci/software/pipelines/espresso-das/espresso-das-pipe-recipes.html}}, version 1.04. We chose the order-by-order format to preserve the original detector binning throughout the coaddition procedure (merged spectra are already rebinned on a fixed wavelength grid). The coaddition was performed with the approach described in \citet{cupani+2016}, equalizing the flux from the different exposures and appropriately weighting the contributions to each bin of the final spectrum. In what follows, unless otherwise specified, all quantities should be considered to be in the rest frame system.

\begin{figure*}
    \centering
    \includegraphics[scale = 0.4]{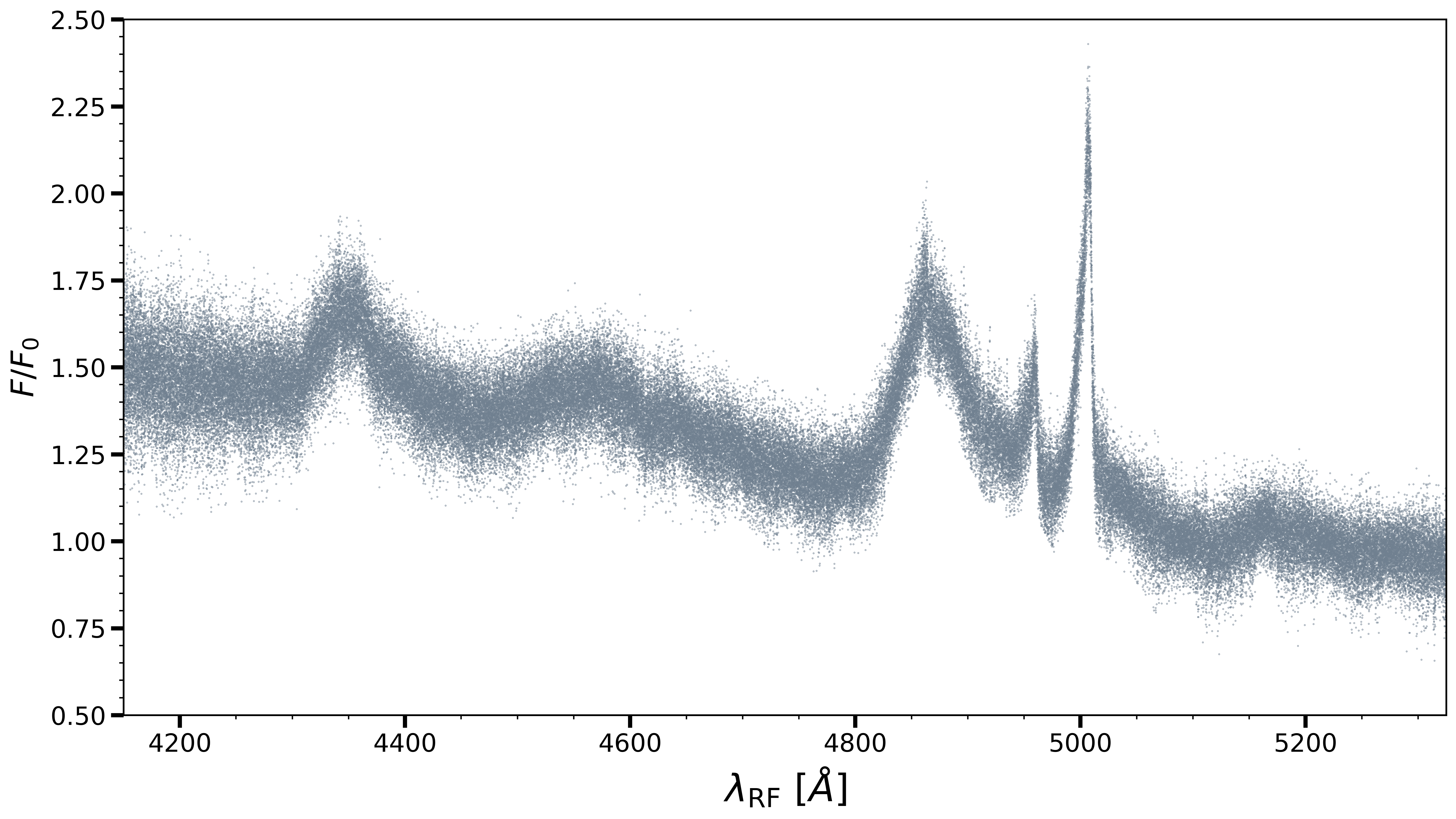}
    \caption{Data considered in the analysis. In the panel, we show the flux normalized to a reference flux, $F_0$, as a function of the rest-frame wavelength of the spectrum of PG~1302-102 at the ESPRESSO native resolution.}
    \label{fig:raw_data}
\end{figure*}

\begin{table*}
	\centering

 \caption{Summary of the model parameters for the multi-Gaussian fit of the spectrum. }
	\label{tab:parameters_gauss}
	{\begin{tabular}{llcccc} 
        \hline\vspace{-0.75em}\\
		Description & Name & Prior range & Model 1 & Model 2\vspace{0.2em}\\
        \hline\vspace{-0.75em}\\
		Broad components offset & $\Delta \mu_{\rm{B}}~ [\rm{km/s}]$ & $[3,3000]$ & $441^{+8}_{-8}$ & $748^{+7}_{-7}$\vspace{0.2em}\\
  		Very broad components offset & $\Delta \mu_{\rm{VB}}~ [\rm{km/s}]$ & $[-30000,30000]$ & $3990^{+90}_{-90}$ & $-$\vspace{0.2em}\\
		H$\beta$ narrow peak flux & $\log_{10}{\rm{A}_{\rm{H}\beta_{\rm{N}}}}$ &$[-3,0.75]$ & $-1.13^{+0.02}_{-0.02}$ & $-1.13^{+0.03}_{-0.03}$ \vspace{0.2em}\\
		H$\beta$ broad peak flux & $\log_{10}{\rm{A}_{\rm{H}\beta_{\rm{B}}}}$ & $[-3,0.75]$ & $-0.363^{+0.002}_{-0.003}$ & $-0.337^{+0.002}_{-0.002}$ \vspace{0.2em}\\
  		H$\beta$ very broad peak flux & $\log_{10}{\rm{A}_{\rm{H}\beta_{\rm{VB}}}}$ & $[-3,0.75]$ & $-0.94^{+0.01}_{-0.01}$ & $-$ \vspace{0.2em}\\
		FWHM broad components & $\log_{10}{\rm{FWHM}_{\rm{B}}}~ [\rm{km/s}]$ & $[3,4.5]$ & $3.599^{+0.002}_{-0.002}$& $3.700^{+0.002}_{-0.002}$\vspace{0.2em}\\
  		FWHM very broad components & $\log_{10}{\rm{FWHM}_{\rm{VB}}}~ [\rm{km/s}]$ & $[3,4.5]$ & $4.10^{+0.01}_{-0.01}$& $-$\vspace{0.2em}\\
  		Outflow components offset & $\Delta\mu_{\rm{O}} ~ [\rm{km/s}]$ & $[-3000,3]$ & $-344^{+4}_{-4}$ & $-335^{+3}_{-3}$\vspace{0.2em}\\
    	  $[\rm{OIII}]$ at $4958\AA$ narrow peak flux & $\log_{10}{\rm{A}_{\rm{[OIII]}_{4958,N}}}$ & $[-3,0.75]$ & $-0.665^{+ 0.004}_{-0.004}$ & $-0.662^{+0.003}_{-0.003}$\vspace{0.2em}\\
		$\rm{[OIII]}$ at $4958\AA$ outflow peak flux & $\log_{10}{\rm{A}_{\rm{[OIII]}_{4958,O}}}$ & $[-3,0.75]$ & $-0.762^{+0.003}_{-0.003}$ & $-0.764^{+0.003}_{-0.003}$ \vspace{0.2em}\\
      	$[\rm{OIII}]$ at $4363\AA$ narrow peak flux & $\log_{10}{\rm{A}_{\rm{[OIII]}_{4363,N}}}$ & $[-3,0.75]$ & $-2.0^{+ 0.2}_{-0.2}$ & $-2.7^{+0.2}_{-0.2}$\vspace{0.2em}\\
       	$\rm{[OIII]}$ at $4363\AA$ outflow peak flux & $\log_{10}{\rm{A}_{\rm{[OIII]}_{4363,O}}}$ & $[-3,0.75]$ & $-1.16^{+0.02}_{-0.02}$ & $-1.03^{+0.01}_{-0.01}$ \vspace{0.2em}\\
		  FWHM outflow components & $\log_{10}{\rm{FWHM}_{\rm{O}}} ~ [\rm{km/s}]$ & $[0,3.5]$ & $2.996^{+0.003}_{-0.003}$ & $3.123^{+0.003}_{-0.003}$  \vspace{0.2em}\\
		FWHM narrow components & $\log_{10}{\rm{FWHM}_{\rm{N}}} ~ [\rm{km/s}]$ & $[0,3]$ & $2.507^{+0.005}_{-0.005}$ & $2.533^{+0.005}_{-0.005}$\vspace{0.2em}\\
		Continuum normalization & $A_{\rm{cont}}$ & $[-10^{-6},30]$ & $5.0^{+0.2}_{-0.2}$ & $0.885^{+0.003}_{-0.003}$ \vspace{0.2em}\\
		Continuum slope parameter & $\alpha_{\rm{cont}}$ & $[-30,30]$ & $-5.7^{+0.2}_{-0.2}$ & $-29.0^{+0.5}_{-0.5}$\vspace{0.2em}\\
  		Continuum slope parameter & $\beta_{\rm{cont}}$ & $[-30,30]$ & $-4.7^{+0.1}_{-0.1}$ & $-20.6^{+0.3}_{-0.4}$\vspace{0.2em}\\
    	Continuum slope parameter & $\gamma_{\rm{cont}}$ & $[-30,30]$ & $-1.8^{+0.1}_{-0.1}$ & $-1.33^{+0.01}_{-0.01}$\vspace{0.2em}\\
        Continuum wavelength break & $\lambda_{\rm{cont}}~[\AA]$ & $[3000,6000]$ & $4310^{+68}_{-51}$ & $5795^{+8}_{-8}$\vspace{0.2em}\\
		  redshift & $z$ & $[0.2,0.3]$ & $0.2786^{+0.0}_{-0.0}$ & $0.2785^{+0.0}_{-0.0}$ \vspace{0.2em}\\
		H$\beta$ outflow peak flux & $\log_{10}{\rm{A}_{\rm{H}\beta_{\rm{O}}}}$ & $[-3,0.75]$ & $-1.16^{+0.02}_{-0.02}$ & $-0.91^{+0.01}_{-0.01}$ \vspace{0.2em}\\
		H$\gamma$ narrow peak flux & $\log_{10}{\rm{A}_{\rm{H}\gamma_{\rm{N}}}}$ & $[-3,0.75]$ & $-1.21^{+0.03}_{-0.03}$ & $-1.25^{+0.03}_{-0.04}$ \vspace{0.2em}\\
		  H$\gamma$ broad peak flux & $\log_{10}{\rm{A}_{\rm{H}\gamma_{\rm{B}}}}$ & $[-3,0.75]$ & $-0.710^{+0.005}_{-0.005}$ & $-0.803^{+0.005}_{-0.006}$ \vspace{0.2em}\\
		H$\gamma$ outflow peak flux & $\log_{10}{\rm{A}_{\rm{H}\gamma_{\rm{O}}}}$ & $[-3,0.75]$ & $-2.1^{+0.2}_{-0.2}$ & $-1.29^{+0.03}_{-0.02} $\vspace{0.2em}\\
		  $\rm{Fe}$ broad peak flux & $\log_{10}{\rm{A}_{\rm{Fe}_{\rm{B}}}}$ & $[-3,0.75]$ & $-0.724^{+0.005}_{-0.005}$ & $-0.783^{+0.002}_{-0.002}$\vspace{0.2em}\\
		$\rm{Fe}$ narrow peak flux & $\log_{10}{\rm{A}_{\rm{Fe}_{\rm{N}}}}$ & $[-3,0.75]$ & $-1.76^{+0.06}_{-0.06}$ & $-1.86^{+0.06}_{-0.08}$\vspace{0.2em}\\
        \hline\vspace{-0.75em}\\
        log evidence & $\log{Z}$ & $-$ & 215929 & 214720 &\vspace{0.2em}\\
        \hline
        
	\end{tabular}
    \tablefoot{From left to right, the columns provide a brief description of each parameter, its reference name as used in this work, the assumed prior range, and the best-fit values with their credibility intervals for two different models. Model 1, differently from Model 2, includes a very broad and, if associated with the H$\beta$, very redshifted Gaussian component. Priors are assumed to be either uniform or log-uniform for each parameter. The last row reports the evidence of the two models.}    }

\end{table*}

\section{Modeling of the spectrum}
\label{sec:modelling}

Here, we describe the specific steps needed to obtain an accurate model of the spectrum of PG~1302-102. More precisely, in \S~\ref{sec:multi_gauss}, we present the scenario of a BLR orbiting in circular motion around a common center of mass of a putative SMBHB. Such a model, in a simplified configuration, can be represented by a Gaussian broad emission profile shifted in wavelength with respect to the rest frame emission of its narrow line. In \S~\ref{sec:disck+spiral}, we propose an alternative disk-like BRL model whose emissivity distribution is distorted by the presence of an azimuthal perturbation. In this scenario, the resulting emission features are not necessarily Gaussian and can trace the presence of bulk motion, possibly associated with the presence of a SMBHB, in the BLR.

\subsection{Multi-Gaussian fit}
\label{sec:multi_gauss}
In this section, we focus on the identification of substructures and asymmetries in the BELs and NELs of the spectrum of PG 1302-102, which may be an indication of the presence of a SMBHB. More specifically, we aim at identifying in the emission lines one or multiple broad components shifted in wavelength that could be associated with the orbital motion of a SMBHB.

In Fig.~\ref{fig:raw_data}, we show the spectrum considered for the analysis. We selected the rest-frame spectral region from $4150\AA$ to $5325\AA$; such choice balances the need to maximize the amount of information extracted from the data with the computational performance of the fit\footnote{Due to the high spectral resolution of ESPRESSO data, we are already considering \fr{$\simeq 138,000$} data points in the fit. Increasing this number further would mainly result in a worsening of computational performance. We note that fitting the current data requires about 5 million likelihood evaluations resulting in a total computational cost for a single parameter estimation of about 40 hours. }. The selected region is large enough to include the $\rm{H}\gamma$ and $\rm{H}\beta$ BELs and the $[\rm{O III}]$ forbidden NELs. Increasing the region further would not have led to significant changes in the extracted best-fit parameters, since it already includes the most prominent emission lines in the wavelength range covered by ESPRESSO for this source.

The full spectrum is modeled as the superposition of different components whose free parameters are constrained by fitting the data with a nested sampling algorithm \citep{Skilling2004}. This approach was chosen since it allows us to measure the Bayesian evidence ($Z$)\footnote{We stress that, comparing the Bayesian evidence of two models is formally the best approach to model selection. Indeed, differently from other indicators (i.e., $\chi^2$, BIC, AIC), the Bayesian evidence automatically accounts for the dimensionality of the parameter space without assuming any shape of the posterior distribution.}, which we use to compare different models and assess the optimal number of components to be included in the fit. More specifically, the search for the optimal model for the spectrum of PG~1302-102 involved a step-by-step process where models of increasing complexity were compared with each other. \fr{This approach automatically prevents overfitting as the Bayesian evidence, being an integral over the whole parameter space, is penalized for models with a larger number of parameters.} The initial model consisted of a power-law continuum along with broad and narrow Gaussian emission lines centered at the expected redshift of the source. From this baseline, we included additional emission components and allowed the already included BELs to be shifted in wavelength. At each stage, we checked that any proposed modification was statistically justified by an increase in the Bayesian evidence. The fiducial model (hereinafter Model 1) was selected based on the highest Bayesian evidence.

Model 1 accounts for a smooth broken power law continuum parameterized as: 

\begin{equation}
    F(\lambda) = A_{\rm{cont}}\left(\frac{\lambda}{\lambda_{\rm{cont}}}\right)^{\alpha_{\rm{cont}}}\left[1+\left(\frac{\lambda}{\lambda_{\rm{cont}}}\right)^{\beta_{\rm{cont}}}\right]^{\gamma_{\rm{cont}}},
    \label{eq:continuum}
\end{equation}

where $A_{\rm{cont}}$ is the normalization, $\lambda_{\rm{cont}}$ is the wavelength break and, $\alpha_{\rm{cont}}$ is the slope at $\lambda \ll \lambda_{\rm{cont}}$, while $\beta_{\rm{cont}}$ and $\gamma_{\rm{cont}}$ define the slope at $\lambda \gg \lambda_{\rm{cont}}$ and ensure a smooth transition at $\lambda\simeq\lambda_{\rm{cont}}$. The modeling of all emission lines (i.e., $\rm{H}\gamma$, $\rm{H}\beta$, and the $[\rm{O III}]$\footnote{[OIII] emission line rest-frame wavelengths: $4363.2\AA$, $4958.9\AA$, and $5006.8\AA$.}  with the Fe contribution discussed separately) includes a narrow Gaussian component with shared parameters for redshift (z) and full width at half maximum (FWHM). These common parameters are applied consistently across the different lines in terms of their velocities.
 Additionally, we included a broad Gaussian emission for the $\rm{H}\beta$ and $\rm{H}\gamma$ whose $\rm{FWHM}_B$ have been assumed to be the same. Moreover, for both emission lines, we accounted for a possible redshift in wavelength (see "Prior range" in Tab.~\ref{tab:parameters_gauss}) with respect to their associated narrow component. This choice was necessary to obtain a reasonably good fit of the spectrum and was also supported by the increase in the Bayesian evidence compared to a model with a fixed centroid. For all [OIII] and Balmer NELs, we included an additional blueshifted Gaussian component, as a single narrow Gaussian emission profile could not adequately fit the data. The additional Gaussians share the same wavelength shift ($\Delta\mu_{O}$) and width ($\rm{FWHM_O}$) for each line. This choice was again driven by the observed increase in the Bayesian evidence when including such additional components. As we shall further discuss in this section and in Sec.~\ref{sec:discussion}, this strongly suggests the  presence of an outflow in the narrow line regions of PG~1302-102. 


The considered region of the spectrum contains a contribution from the Fe emission lines \citep[][]{Boroson1992,Veron_cetty_2004,Tsuzuki2006}{}{}. We accounted for the Fe contribution following \cite{Calderone2017}. More precisely, we consider all the BELs and NELs listed in \cite{Veron_cetty_2004} and, for each of these lines we included a Gaussian component. All the Fe NELs have the same redshift and FWHM as the narrow components of the $\rm{H}\gamma$, $\rm{H}\beta$, and $[\rm{O III}]$. Similarly, we assume the Fe BELs to have the same FWHM ($\rm{FWHM_B}$) and shift in wavelength ($\Delta\mu_{\rm{B}}$) as the broad components of the $\rm{H}\gamma$ and $\rm{H}\beta$.

Each Gaussian has a different amplitude parameter $\rm{A}_j$ (see Tab.~\ref{tab:parameters_gauss} for the definition of the different subscripts) defined as the peak of each Gaussian relative to a common reference flux\footnote{The spectrum considered in the analysis is normalized to a reference flux $\rm{F}_0 = 2.96 \times 10^{-15} \rm{erg/s/cm^2/\AA}$ chosen to have a flux close to 1 at $\lambda\simeq5100\AA$.}. The amplitude ratio between the $[\rm{OIII}]\lambda4959\AA$ and the $[\rm{OIII}]\lambda5007\AA$ is assumed to be 1:3 \citep[][]{Osterbrock2006}, while the ratio of the individual Fe lines is taken from \cite{Veron_cetty_2004}. We note that our approach is one of the most general in treating the Fe contribution in type I AGNs; we allow for independent broad and narrow emissions across all expected wavelengths without constraining the fit to any specific template. To check if our results depend on the specific method adopted to model the Fe contribution, we compared it against other approaches \citep[e.g.,][]{Boroson1992,Tsuzuki2006,Marziani2009}{}{} where the Fe contribution is modeled by applying a Gaussian kernel on a reference template. We find that the considered templates were statistically unfavored and that the estimates on the other parameters were not significantly affected by the specific method adopted. 

Finally,  Model 1 includes an additional ``very'' broad component that is also significantly redshifted if it is associated with the H$\beta$ emission line (see the discussion below). Such component is described through a Gaussian profile with its own amplitude ($\rm{A}_{\rm{VB}}$), wavelength shift with respect to the H$\beta$ narrow line ($\Delta\mu_{\rm{VB}}$), and FWHM ($\rm{FWHM}_{\rm{VB}}$) parameters. As we discuss below, the same component is not statistically required, and therefore not included in Model 1 for the H$\gamma$.  Model 2 is a replica of Model 1, but for the fact that it does not include the additional very broad component.

We report in Tab.~\ref{tab:parameters_gauss} a summary of the model parameters, their prior\footnote{We assumed either uniform or log-uniform priors for all the parameters.} and the best-fit estimates from Model 1 and Model 2. The last row of Tab.~\ref{tab:parameters_gauss} reports the evidence of the two models that points toward a decisive preference \citep[according to][]{Jeffreys1939} for Model 1 against Model 2 ($\Delta\log{Z}=1209$). The most significant differences among the estimated parameters of the two models regard the continuum and the broad component of the Balmer line. The differences in the continuum parameters might be explained as a consequence of the relatively small wavelength range considered. Although the parameters defining the continuum of Model 1 and Model 2 are quite different, still, the predicted continuum flux is almost indistinguishable between the two cases in the considered wavelength range. On the other hand, the difference in the broad Balmer lines is mainly due to the inclusion of the very broad Gaussian component in Model 1, which reduces $\Delta\mu_{\rm{B}}$ (from $\simeq12\AA$ to $\simeq7\AA$ in the case of the H$\beta$) compared to Model 2. Still, in both cases, the presence of asymmetry towards red wavelengths remains evident for the Balmer lines.

\begin{figure*}
    \centering
    \includegraphics[width=1.0\textwidth,height=0.75\textwidth]{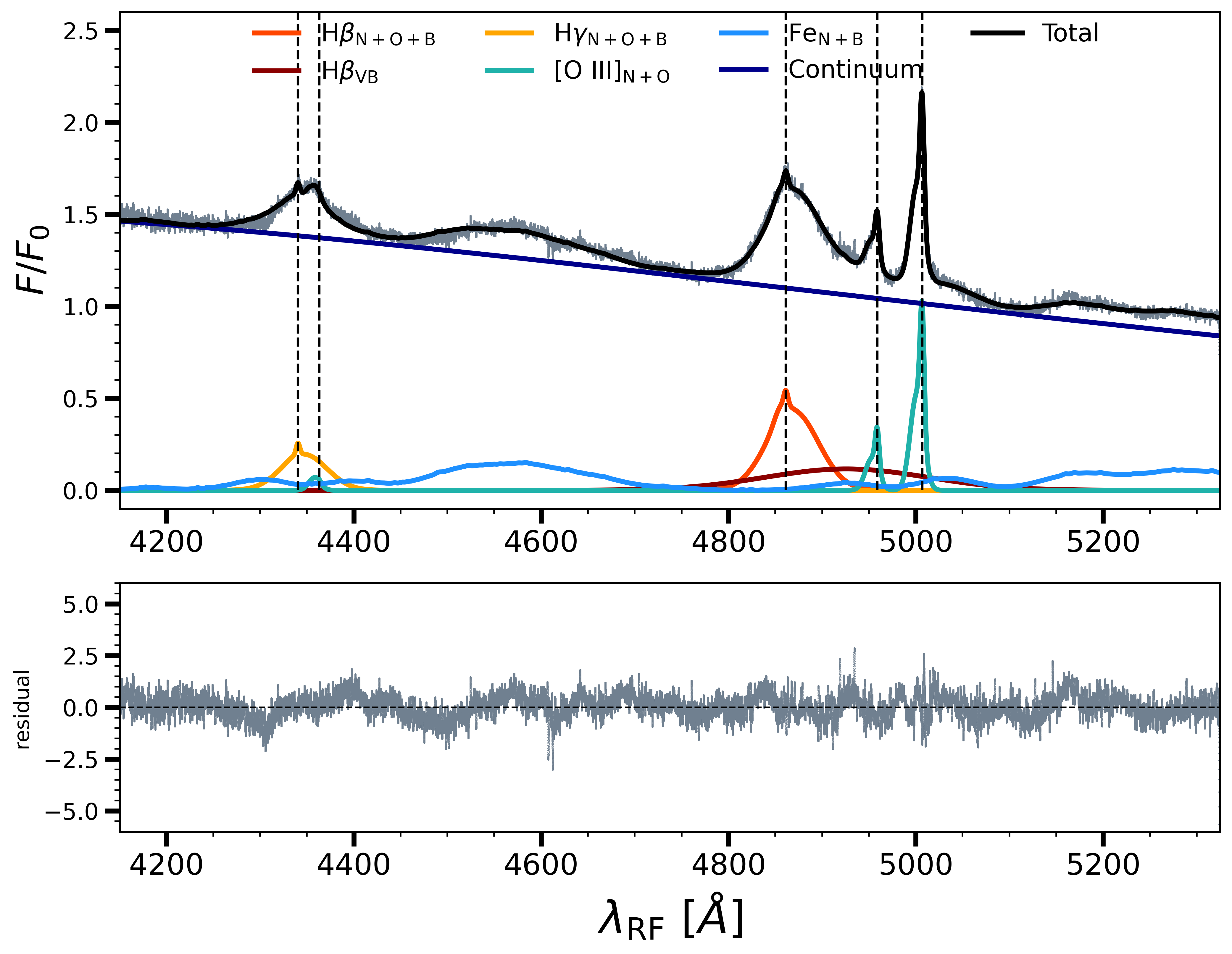}
    \caption{Best-fit result of Model 1 on the ESPRESSO data. The top panel shows the data in gray smoothed with a Gaussian Kernel ($\sigma \simeq 0.07\AA$), the best-fit model (black line), and all the emission components identified by different colors: $\rm{H\gamma}$ in gold, $\rm{H\beta}$ in red, $[\rm{OIII}]$ in cyan, Fe in light-blue, very broad $\rm{H\beta}$ in brown, and the continuum in dark blue. The vertical dashed lines indicate, from left to right, the rest-frame emission wavelength of the $\rm{H\gamma}$, $[\rm{OIII}]\lambda4363.2\AA$, $\rm{H\beta}$, and $[\rm{OIII}]\lambda4958.9\AA,\lambda5006.8\AA$. The bottom panel represents the residuals defined as (data-model)/error, and the horizontal dashed black line at the zero level of residuals is plotted to guide the eyes.}
    \label{fig:bf_fiducial}
\end{figure*}

Fig.~\ref{fig:bf_fiducial} shows the result of the fit of Model 1 against the ESPRESSO data (the results obtained from Model 2 are reported in appendix~\ref{App:model2}). The top panel shows the data in gray with a Gaussian smoothing ($\sigma\simeq0.07\AA$) and the best-fit model with a black line. The contributions of the continuum and each individual emission line to the overall spectrum are highlighted in different colors. The bottom panel shows the residuals defined as the difference between the model and the data divided by the error where the typical error on the data is $\simeq 0.045$ in units of $F_0$. Overall, Model 1 reproduces the data well, as can be deduced by the lack of strong systematics in the residuals and the agreement observed between the smoothed spectra.


The superposition of the different Gaussian contributions results in emission profiles that deviate from a symmetric shape. More specifically, the BELs of the $\rm{H}\gamma$, $\rm{H}\beta$ are clearly redshifted (see $\Delta\mu_{\rm{B}}\simeq440~\rm{km/s}$ in Tab.~\ref{tab:parameters_gauss}). \fr{Such an asymmetric shape of the emission profile was already observed} in \citealt{Graham2015Nat} although only for the $\rm{H}\beta$ and the $\rm{H}\alpha$\footnote{The authors report the presence of a redshifted tail also in the $\rm{H}\alpha$. Since that line does not fall in the observed wavelength range we cannot confirm its presence.} and not for the $\rm{H}\gamma$. \fr{In this work, given the high resolution of the ESPRESSO data, we confirm the presence of such a redshifted component and characterize its features with higher precision}. Moreover, we also detect, for the first time, the presence of such redshifted asymmetry in the H$\gamma$ emission line. Given our results and the considerations made in \citealt{Graham2015Nat} about the presence of a similar redshifted component in the H$\alpha$, we argue that such behavior could be common among all the Balmer lines. 

Our fiducial model indicates the presence of an additional broad ($\rm{FWHM_{VB}}\simeq12600\rm{km/s}$) Gaussian component that appears to be very redshifted ($\Delta\mu_{\rm{VB}}\simeq4000 \rm{km/s}$) if associated to the H$\beta$ emission. The improvements in the modeling when including such component are quite significant as can be deduced by comparing the residual between Fig.~\ref{fig:bf_fiducial} and Fig.~\ref{fig:bf_no_outflow} in the region comprising the H$\beta$ and $\rm{[OIII]}$ doublet and by the increase in the Bayesian evidence. This is the first time such a broad and extremely redshifted component has been detected in PG~1302-102. However, it seems that a non-negligible fraction of quasars, with a spectrum similar to that of PG~1302-102 (i.e., asymmetric and redshifted H$\beta$ emission line), shows the presence of such feature \citep[e.g.,][]{Sulentic2002,Marziani2009}. The physical origins of this peculiar component are still debated (see Sec.~\ref{sec:discussion} for a more detailed discussion), and its inclusion is usually justified on empirical grounds, as it significantly improves the fit residuals. If such a component is connected to a process in the BLR\fr{,} it might be reasonable to expect its presence to be ubiquitous in all the Balmer lines. We attempted fitting for an additional very broad and very redshifted component also for the H$\gamma$ emission line without finding any increase in the Bayesian evidence and, therefore rejecting the detection of such a component in the H$\gamma$\footnote{Even fixing the amplitude ratio $A_{\rm{H\beta_B}}/A_{\rm{H\beta_{VB}}}$ to be the same of $A_{\rm{H\gamma}}/A_{\rm{H\gamma_{VB}}}$ we did not observe any increase in the model evidence.}.

The additional Gaussian component included in the modeling of $[\rm{OIII}]$, H$\gamma$, and H$\beta$ NELs is shifted toward bluer wavelengths, ($\Delta\mu_{\rm{O}}\simeq-340 \rm{km/s}$) and is relatively broad ($\rm{FWHM_O}\simeq1260~\rm{km/s}$). Given the observed properties, we interpret it as the presence of an ionized gas outflow affecting the NLR of PG~1302-102. We stress that, this is the first detection of a possible outflow component in PG~1302-102, which has only been made possible thanks to the high resolution of ESPRESSO.

From the BELs parameter (i.e., $\rm{FWHM_B}$), assuming the BLR being virialized and the broad component of the Balmer lines tracing the gravitational potential dominated by a SMBH, we can estimate its mass ($M_{\rm{BH}}$) through single epoch method \citep[see Tab.~4 of][]{Shen2024}{}{}:

\begin{equation}
\begin{split}
    \log_{10}{\left(\frac{M_{\rm{BH}}}{M_{\sun}}\right)} = 0.85+&0.5\log_{10}\left({\frac{L_{5100}}{10^{44} \rm{erg/s}}}\right) +\\ &2.0\log_{10}{\left(\frac{\rm{FWHM}}{\rm{km/s}}\right)},
\end{split}
\label{eq:SEM}
\end{equation}

\noindent where $L_{5100}$ is the luminosity at $\lambda=5100\AA$ and the $\rm{FWHM}$ is the full width at half maximum of the broad $\rm{H}\beta$ emission line. If relying on the \fr{estimates} from Model 1, using $L_{5100}=(3.66 \pm 0.03)\times 10^{45}\rm{erg/s}$ and $\rm{FWHM}=\rm{FWHM_B} = 3972 \pm 18\rm{km/s}$, we get $\log_{10}{\rm{M_{\rm{BH}}}/\rm{M_{\sun}}}= 8.83\pm0.46$, where most of the error comes from propagating the scatter of Eq.~\ref{eq:SEM}. This number is in line with previously reported estimation \citep[][]{Graham2015Nat}.



\subsection{Single disky broad-line region with a spiral perturbation}
\label{sec:disck+spiral}
BLRs are often considered spherically symmetric and virialized. However, multiple observational results, mostly from reverberation mapping campaigns, have proven this is not always the case \citep[][]{Eracleous1995,Bentz2010,Grier2017}. This is especially true when emission lines deviate from pure Gaussianity \citep[][]{Zastrocky2024} as this is often tracing the presence of sub-structures in the BLR clouds distribution or deviations from virial equilibrium \citep[][]{Pancoast2014}. Even though the structure of the BLR has been typically investigated with velocity-resolved reverberation mapping, here we follow an approach closer to that presented in \citealt{Raimundo2020} on single epoch observations. Similarly to what is done in that work, we employ a nested sampling approach for the parameter estimation\fr{;} however, instead of focusing on the BLR kinematics, we consider asymmetric deviation in its emissivity. Our semi-analytic model is based on \citealt{Storchi-Bergmann} where asymmetric emission lines are modeled by introducing a spiral perturbation in a disk-like BLR rotating in circular Keplerian motion around its SMBH. Such kind of perturbations are expected to arise in the self-gravitating part of accretion disks \citep[][]{Wang2022}.

In what follows, we summarize the main assumptions of the model ``spiral model'' hereinafter) that is described in detail in Sottocorno et al. (in prep.). The BLR is assumed to be on a razor-thin disk-like configuration whose axial symmetry is broken due to the presence of a spiral perturbation. Under these assumptions, the BLR emissivity profile reads:

<
\begin{equation}
\begin{split}
 \epsilon (\xi, \phi) = \xi^{-1}&\cdot\exp\left[-\frac{(\xi-\xi_c)^2}{2\sigma_{\xi_c}^2}\right]\Biggl\{ 1 +\\ &\frac{A}{2}\exp\left[-\frac{4\log{2}}{\delta^2}(\phi - \psi_0)^2\right]+\\
 & \frac{A}{2}\exp\left[-\frac{4\log{2}}{\delta^2}(2\pi - \phi + \psi_0)^2\right]\Biggr \},
\end{split}
\label{eq:emissivity}
\end{equation}

\noindent where $\xi$ is the cylindrical radius normalized to the gravitational radius $R_g=GM_{\rm{BH}}/c^2$, and $\phi$ is the azimuthal angle. $\xi_c$ and $\sigma_{\xi_c}$ are the radial position and the extent of the bulk of the emissivity profile respectively, while $A$ and $\delta$ define the relative intensity of the spiral perturbation with respect to the underground Gaussian emissivity and the azimuthal width of the spiral, respectively. Finally, $\phi - \psi_0$ denotes the azimuthal distance from the ridge of the spiral arm, where $\psi_0 = \phi_0 + \log_{10}{(\xi/\xi_{sp})}/\tan{p}$. Here $\phi_0$ is the initial angular position of the spiral at the innermost radius of the spiral $\xi_{sp}$, while $p$  is the spiral pitch angle. 

The emissivity profile of Eq.~\ref{eq:emissivity} differs from that assumed by \citealt{storchi03} for a Gaussian multiplicative factor in the radial dependence of the emissivity. This choice ensures that the bulk of the BLR emissivity is localized close to the radius predicted by the luminosity-radius relation \citep[][]{bentz09b,Shen2024}. 

The emission line profiles of the BLR are the emissivity-weighted sum of the contribution coming from each resolution element of the BLR scaled by a luminosity normalization parameter $L_{\rm{H}\beta}$ (for the H$\beta$) and $L_{\rm{H}\gamma}$ (for the H$\gamma$). We assume each resolution element to emit as a Gaussian centered at the $\rm{H}\beta$ (or H$\gamma$) Doppler-shifted wavelength and whose width is regulated by a broadening parameter, $\sigma_\lambda$. The Doppler shift of each resolution element is computed by assuming the BLR is inclined by an angle, $i$, with respect to line-of-sight.

A summary of the model parameters together with the assumed priors and the best-fit parameters is reported in Tab.~\ref{tab:parameters_spiral_model}. We assumed either uniform or log-uniform priors for most of the parameters, except for the inclination angle which follows a uniform distribution for $\sin{i}$. We allowed a maximum inclination angle of $\pi/4$ since for a higher inclination angle, we do not expect to observe BELs due to the obscuration from the dusty torus. Moreover, we assume the BLR extending between 200 and 18000 gravitational radii, which, in the case of a $10^{8.8}~M_{\odot}$ BH\fr{,} becomes from $\simeq 0.006~\rm{pc}$ to $\simeq 0.58~\rm{pc}$. {This number must be considered as an order of magnitude estimate of the BLR size since the gravitational radius is affected by the error propagation on the BH mass. Still, since the spiral model does not explicitly depend on the BH mass, the results regarding the appearance of PG~1302-102 BLR will remain unchanged.}. Finally, since the H$\beta$ and H$\gamma$ emission line might not necessarily be localized at the same distance from the SMBH and subject to the same dynamics \citep[][]{Kuhn2024}, we included different radial lengths, widths, and broadening parameters for the two lines. 

The spiral model presented here does not account for anything else besides a broad emission component at a chosen reference wavelength (the H$\beta$ and H$\gamma$ in our case). For this reason, in order to apply such a model directly to the spectrum presented in Fig.~\ref{fig:raw_data}, we must account for all the sources of emission that are not associated with a broad component (``background model'' hereinafter). Since the spiral model has a higher computational cost, to limit the number of iterations before reaching convergence we fixed the background model to the best fit obtained in Model 1 excluding the broad Gaussian contribution associated with the H$\beta$, H$\gamma$, and the very broad and redshifted component. Even with this approximation the overall parameter estimation required more than a week\footnote{On an 8-core machine equipped with 13th Gen Intel(R) Core(TM) i7-1355U.} before reaching convergence. Therefore, in order to obtain a complete parameter estimation without excluding any data\footnote{A possibility could have been to focus on the H$\beta$ only reducing the analyzed wavelength range and the overall computational cost of the parameter estimation.}in a reasonable amount of time, we followed the approach outlined in \citealt{Rigamonti2022,Rigamonti2023}, implementing a GPU parallelization strategy for a much faster (few hundred times) computation of the likelihood function. 

\begin{figure*}
\includegraphics[width=1.0\textwidth,height=0.75\textwidth]{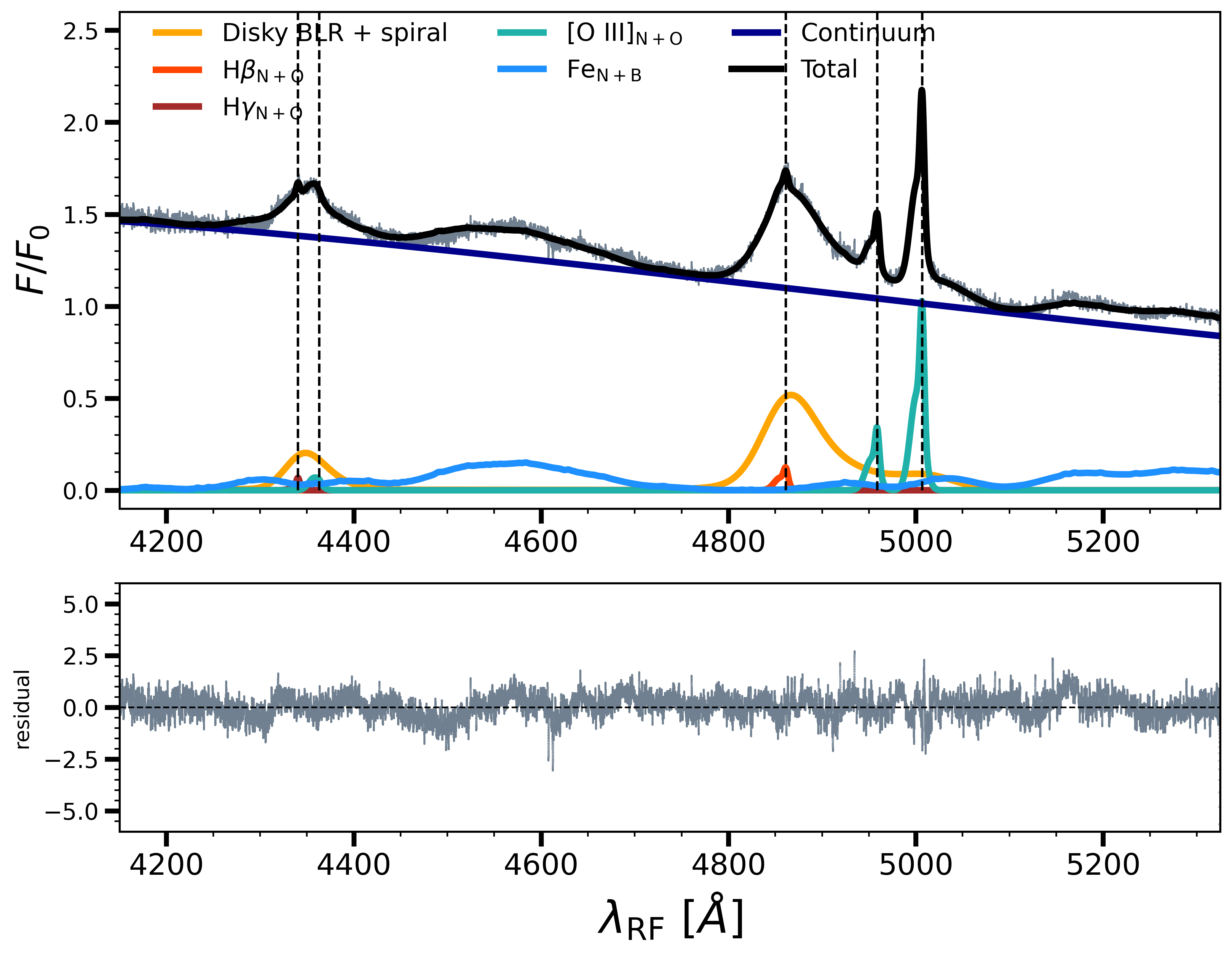}
\caption{Best-fit result of the spiral model on the ESPRESSO data. The top panel shows the data in gray smoothed with a Gaussian Kernel ($\sigma \simeq 0.07\AA$), the best-fit model in black, the spiral model in gold, and all the emission components of the background model identified by different colors: $\rm{H\gamma}$ in brown, $\rm{H\beta}$ in red, $[\rm{OIII}]$ in cyan, Fe in light-blue and the continuum in dark blue. The vertical dashed lines indicate, from left to right, the rest-frame emission wavelength of the $\rm{H\gamma}$, $[\rm{OIII}]\lambda4363.2\AA$, $\rm{H\beta}$, and $[\rm{OIII}]\lambda4958.9\AA,\lambda5006.8\AA$. The bottom panel represents the residuals defined as (data-model)/error, the horizontal dashed black line at the zero level of residuals is plotted to guide the eyes.}
\label{fig:Best-fit_spiral_model}
\end{figure*}

The top panel of Fig.~\ref{fig:Best-fit_spiral_model} shows the best-fit spiral model in black (see Tab.~\ref{tab:parameters_spiral_model} for the estimated parameters) on top of the ESPRESSO data in gray. The gold line refers to the broad emission from the spiral BLR, while all the other colors refer to the components of the background model that have been taken from Model 1. The bottom panel represents the residuals defined as (data-model)/error. Compared to Model 1, the spiral model provides a better fit to the data, reducing the residuals both in the region dominated by the H$\beta$ and the H$\gamma$. Notably, the spiral model is able to account for the excess at $\lambda\simeq5000\AA$ giving a physical interpretation to the very broad and very redshifted component included in Model 1. The same model, without changing the overall geometrical structure\footnote{We note that $i$, $\rm{A}$, $\delta$, $\phi_0$, and $p$ are the same for the H$\beta$ and H$\gamma$.} of the BLR, improved the fit of the H$\gamma$ emission (see Sec.~\ref{sec:discussion} for a more detailed discussion). Notably, even though we did not impose any prior relating the scale radius, width, and broadening of the H$\beta$ and H$\gamma$ we found quite similar parameters. This suggests that the two emission regions, and possibly all the Balmer lines, share similar dynamics ($\rm{FWHM}_{\rm{H}\beta}\simeq\rm{FWHM}_{\rm{H}\gamma}$) and are spatially coherent ($\xi_{c,\rm{H}\beta}\simeq\xi_{c,\rm{H}\gamma}$), with the H$\gamma$ emission region being slightly more compact than the H$\beta$ ($\sigma_{\xi_{c,\rm{H}\beta}}\gtrsim\sigma_{\xi_{c,\rm{H}\gamma}}$). 

\begin{figure*}
    \centering
    \includegraphics[width=1.0\textwidth,height=0.75\textwidth]{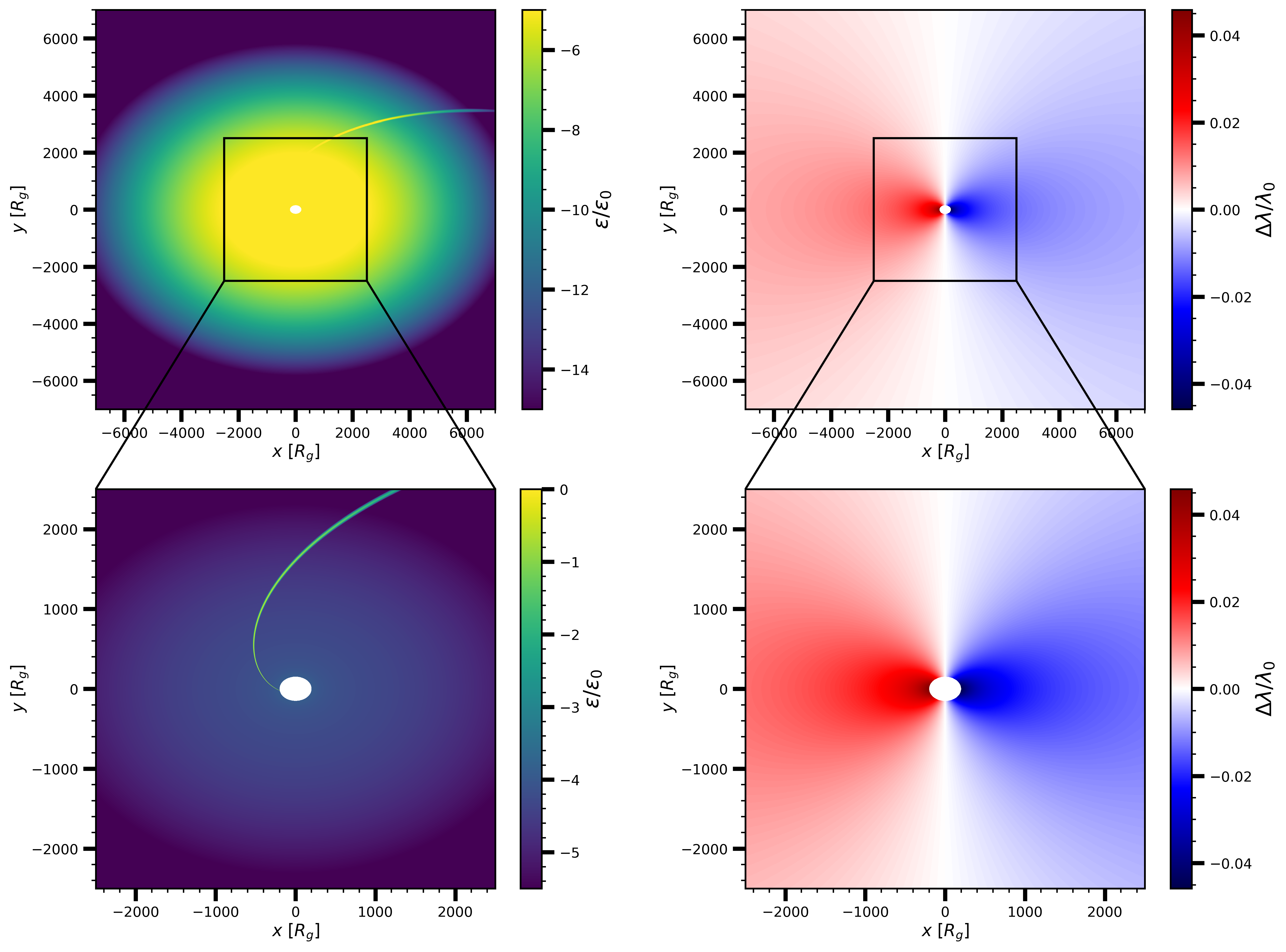}
    \caption{Projected H$\beta$ emission and Doppler shift of the spiral BLR best-fit. The top left panel is the projected BLR emissivity normalized to the maximum emissivity for the H$\beta$, while the right top panel represents the Doppler shift of each BLR element. The bottom panels show a zoom of the emissivity (left) and Doppler shift (right) on the central region of the BLR.}
    \label{fig:2D_projected_emissivity}
\end{figure*}

The specific features of the resulting emission profile can be better understood by inspecting Fig.~\ref{fig:2D_projected_emissivity}, where we show the projected emissivity distribution of the H$\beta$ (see Fig.~\ref{fig:2D_projected_emissivity_Hgamma} for the H$\gamma$) and the map of the BLR Doppler shift. The left column shows the projected emissivity profile together with a zoomed version in the central region. The emissivity steeply decreases after the first 1000 gravitational radii as a consequence of the exponential cut-off and the $\xi^{-1}$ decline. Interestingly, the spiral perturbation is extremely thin in terms of azimuthal width ($\delta = 0.0085$) although very intense ($\log_{10}{\rm{A}}=4.22$) suggesting a strong perturbation of the BLR compared to the underlying disk. The right side of the plot shows the Doppler shift of each BLR element, which, as is expected, decreases following the Keplerian law. The spiral perturbation encompasses the region of the BLR that has a positive Doppler shift, explaining why the emission profile is redshifted. The innermost region, characterized by the highest Doppler factor but lower flux\footnote{The emissivity is higher but the area is smaller.}, is responsible for modeling the excess observer at $\lambda \simeq 5000\AA$. Conversely, the emissivity at larger radii, where the Doppler shift is less significant and the flux emitted from the perturbation becomes less important compared to that from the BLR disk, accounts for the shape of the emission around the bulk of the line (i.e., $\lambda \simeq 4861\AA$).

\begin{table*}
	\centering
	\caption{Summary of the spiral model parameters.}
	\label{tab:parameters_spiral_model}
    {\begin{tabular}{llccc} 
        \hline\vspace{-0.75em}\\
		Description & Name & Prior range & Spiral model\vspace{0.2em}\\
        \hline\vspace{-0.75em}\\
		Inclination & $i~ [\rm{rad}]$ & $[0, \pi/4]$ & $0.71^{+0.01}_{-0.01}$\vspace{0.2em}\\
		Spiral contrast & $\log_{10}{\rm{A}}$ &$[-2,4.5]$ & $4.22^{+0.03}_{-0.03}$ \vspace{0.2em}\\
		Spiral width & $\delta~ [\rm{rad}]$ & $[0, 2\pi]$ & $0.0085^{+0.0004}_{-0.0004}$ \vspace{0.2em}\\
		Spiral azimuthal origin & $\phi_0~ [\rm{rad}]$ & $[0, 2\pi]$ & $1.43^{+0.03}_{-0.001}$\vspace{0.2em}\\
  		Pitch angle & $p~ [\rm{rad}]$ & $[0, \pi]$ & $0.540^{+0.001}_{-0.001}$ \vspace{0.2em}\\
    	  H$\beta$ broadening parameter & $\log_{10}{\rm{FWHM}_{\rm{H\beta}}}~[\AA]$ & $[2,4]$ & $3.573^{+ 0.003}_{-0.003}$ \vspace{0.2em}\\
        H$\gamma$ broadening parameter & $\log_{10}{\rm{FWHM}_{\rm{H\gamma}}}~[\AA]$ & $[2,4]$ & $3.472^{+ 0.005}_{-0.005}$ \vspace{0.2em}\\
		H$\beta$ scale radius & $\log_{10}{\xi_{c,\rm{H}\beta}}~[\rm{R_g}]$ & $[2.3,4.3]$ & $3.13^{+0.1}_{-0.1}$ \vspace{0.2em}\\
  		H$\gamma$ scale radius & $\log_{10}{\xi_{c,\rm{H}\gamma}}~[\rm{R_g}]$ & $[2.3,4.3]$ & $3.190^{+0.002}_{-0.002}$ \vspace{0.2em}\\
		  H$\beta$ radial width & $\log_{10}{\sigma_{\xi_{c,\rm{H}\beta}}}~[\rm{R_g}]$ & $[2.3,4.3]$ & $2.98^{+0.01}_{-0.01}$ & \vspace{0.2em}\\
		  H$\gamma$ radial width & $\log_{10}{\sigma_{\xi_{c,\rm{H}\gamma}}}~[\rm{R_g}]$ & $[2.3,4.3]$ & $2.60^{+0.01}_{-0.01}$ & \vspace{0.2em}\\
		  H$\beta$ luminosity & $\log_{10}{L_{\rm{H}\beta}}~[L_0]$ & $[-2,2]$ & $-0.21^{+0.02}_{-0.03}$ & \vspace{0.2em}\\
		  H$\gamma$ luminosity & $\log_{10}{L_{\rm{H}\gamma}}~[L_0]$ & $[-2,2]$ & $-0.88^{+0.02}_{-0.03}$ & \vspace{0.2em}\\
        \hline\vspace{-0.75em}\\
        log evidence & $\log{Z}$ & - & 206941 \vspace{0.2em}\\
        \hline\vspace{-0.75em}\\  
	\end{tabular}
    \tablefoot{From left to right, the columns detail a brief description of each parameter, its reference name as used in this work, the assumed prior range, and the best-fit values with their statistical errors. Priors are assumed to be either uniform or log-uniform for each parameter, except for the inclination angle, for which we assumed a uniform distribution on the solid angle.}}
\end{table*}

\section{Discussion}
\label{sec:discussion}

The ESPRESSO spectrum of PG~1302-102 clearly revealed the presence of asymmetric emission profiles (in this case skewed toward redder wavelengths) for the broad component of the Balmer lines. This feature was already detected in previous works \citep[e.g.,][]{Boroson1992,Jackson1992,Graham2015Nat} although at a much lower spectral resolution compared to the ESPRESSO data presented here. Also, we detected an excess at $\lambda\simeq5000\AA$ possibly associated with a very redshifted and very broad component originating from the H$\beta$.

The presence of asymmetric BELs has often been associated with the existence of a SMBHB; this hypothesis is particularly intriguing in the case of PG~1302-102, which has already been identified as a candidate SMBHB due to the modulation in its optical light curve \citep{Graham2015Nat}. As was discussed in Sec.~\ref{sec:intro}, the asymmetric emission profile of a SMBHB can result from the orbital motion of two black holes. This occurs when the separation between the black holes is large enough for at least one to retain its own BLR. To satisfy this condition the BLR should not cross the Hill radius of the black hole\footnote{We note that in the case of a disk-like BLR, the radius decreases by a factor of $\simeq 4 - 5$ due to the lack of stability in circular orbits outside of it \cite{runnoe15size}.}. The obvious way to test this hypothesis would be to observe, on an orbital timescale \citep[i.e., $\simeq 5.5\pm0.15 \rm{yr}$ for PG~1302-102,][]{Zhu2020}, the expected Doppler drift in the BEL profiles \citep[][]{popovic2012}. Optical spectra of PG~1302-102 taken at different times are already available \citep[][]{Boroson1992,Graham2015Nat} possibly allowing us to check if a wavelength shift over 5 years is present. However, the spectrum presented in \cite{Graham2015Nat} was taken in April 2014, while ours was during March 2018, giving a time difference too close to the binary orbital period. Similarly, the spectrum presented in \cite{Boroson1992} was taken in April 1990, which, assuming the 5.5 year period of the binary, would be close to exactly 5 periods before our observation. Also, 18 optical spectra of PG~1302-102 were taken from January to June 1990 by \cite{Jackson1992}; even in that case no variability was observed in the H$\beta$ emission line. Still, we have analyzed the spectra presented in \cite{Boroson1992} (the only one publicly available) without finding statistically meaningful differences in the emission profile. The error on the estimated $\Delta \mu_{\rm{B}}$ was large enough to be consistent both with the presence of the Doppler shift reported in Tab.~\ref{tab:parameters_gauss} or with zero shift. We do not exclude this behavior from being caused by the low resolution and low S/N of the data.

In the case of PG~1302-102, it would be tempting to compare the periodicity of the light curve \citep[i.e., $\simeq 5.5\pm0.15~ \rm{yr}$,][]{Zhu2020} with the observed shift in the Balmer BELs (see Fig.~\ref{fig:bf_fiducial}). In the case of the less redshifted component (i.e., $\Delta\mu_{\rm{B}}\simeq450~\rm{km/s}$), assuming circular motion and a binary of mass $10^{8.8}~M_\sun$, we obtain an upper limit for the separation of the two BHs of $\simeq 14~ \rm{pc}$\footnote{Since we do not know the inclination angle and the phase of the putative binary any radius smaller than $14~ \rm{pc}$ cannot be rejected.}. This is not inconsistent with the $\simeq$ 5.5-year period observed in the light curve \citep[][]{Zhu2020}, which, still assuming circular motion, corresponds to a separation of $R \simeq 0.015~ \rm{pc}$ \citep[see Ep.~1 of][]{Yu2001}. However, it must be said that even excluding any kind of obscuration, only few specific combinations of the binary inclination ($i$) and azimuthal ($\phi$) angle result in the right deprojection of the velocity to obtain the $5.5~\rm{yr}$ period. Either the plane of the binary is close to face-on or the line joining the two black holes is parallel to the line of sight. By assuming a uniform distribution for $\phi$ between $-\pi/2$ and $\pi/2$  and a distribution uniform in $\sin{i}$ between $0$ and $\pi/4$ we can give a rough estimate of the probability of obtaining the right deprojection of the velocity to get the $5.5~\rm{yr}$ period of $\simeq 4\%$\footnote{This estimate takes into account the error on the period of the binary ($\simeq 0.15\rm{yr}$) and the error of the black hole mass ($\simeq 0.5\rm{dex}$). If we had assumed to know precisely the black hole mass, the probability would have decreased to $\simeq0.1\%$.  }. Therefore, although the data do not completely rule out the hypothesis of two distinct BLRs producing the observed BEL asymmetry, this scenario is very unlikely. 

Our alternative explanation for the asymmetric emission of the Balmer BEL invokes the presence of a perturbation that breaks the axis symmetry in the disk of a single BLR, which could either surround a single SMBH or a binary SMBH system. Following \cite{Storchi-Bergmann}, we modeled the perturbation with a spiral overdensity. Our analysis indicates such model being statistically preferred\footnote{We note that we cannot directly compare the evidence presented in Tab.~\ref{tab:parameters_gauss} with the evidence estimated in Sec.~\ref{sec:disck+spiral} since the two models have been optimized on different slices of the data.} over a model with a single Gaussian component for both the BELs ($\Delta\log{\rm{Z}}=13437$), a model with a single Gaussian for the H$\gamma$ and two Gaussians for the H$\beta$ ($\Delta\log{\rm{Z}}=1080$) and two Gaussians for both lines ($\Delta\log{\rm{Z}}=6147$).

We expect multiple kinds of perturbations, not necessarily spiral in nature, arising from different physical processes in BLRs. The spiral model is just a simple way to approximate asymmetric emission due to asymmetric emissivity distribution. Indeed, we are not interested in discussing the specific parameters of the spiral model determined in the fit. Instead, we want to stress that even a quite simple toy model is capable of characterizing (still with statistical significance) distorted emission lines. Nevertheless, the presence of spirals in BLR disks seems not to be very uncommon. Theoretical studies \citep[][]{Wang2022} suggest that non-axisymmetric perturbations (spiral structures) may inevitably grow in the self-gravitating parts of AGN accretion disks. These regions are spatially overlapped with BLRs. It is therefore reasonable to believe that inhomogeneities in the self-gravitating parts of accreting disks might be the origin of the spiral arms in disk-like BLRs. Such hypotheses are also supported by the growing number of reverberation mapping velocity-resolved delay analyses. In such studies there are evidences of the common presence of disk-like BLR consistent either with Keplerian motion, inflow, outflow, or with a complex configuration \citep[][]{Bentz2010,Grier2012,Du2018,Zastrocky2024}. 
We also observe that the expected BLR size inferred from the luminosity-radius relation \citep[][]{bentz09b,Shen2024,Wang2024} is a few times larger than the predicted separation of PG~1302-102 ($\simeq0.015~\rm{pc}$). This implies that the individual BLRs of the two SMBHs in PG~1302-102, if really present, should have already settled in a single BLR. However, given the large uncertainties on the BLR size estimates (typically on the order of 0.3 dex) it might be possible that the two BLRs are still in the process of merging and, therefore have been disturbed as a consequence of the passage of two SMBHs. This scenario gives a complementary explanation to that of gravitational instabilities.



We note also that if the variability of PG~1302-102 is due to Doppler boosting, as in the model presented in \citealt{DHS15}, observable effects should be evident by comparing multiple spectra taken at different times \citep[][Bertassi et al. in prep.]{Ji2021}. Indeed, we expect variability in BELs to be different depending on the geometry of the BLR (disturbed or not) and on the source of variability itself (Doppler-boost vs. red noise). \cite{Song2021} analyzed a few spectra taken at different times of PG~1302-102 containing multiple BELs (Ly$\alpha$, $\rm{CIV}$ and $\rm{CIII}$) without observing variability consistent with the Doppler boosting scenario unless assuming strong misalignment between the BLR and the plane of the binary.\footnote{If the binary plane is perpendicular to the BLR, then, in the reference frame of the BLR, the Doppler effect should be smaller. However, given the extended size of the BLR and the uncertainties on its geometry, it is not clear what the net effect would be on the variability of broad lines.}However, their analysis was limited to a few good-quality spectra taken at different times, a factor that could have affected their findings.

Consistently with the results from RM campaigns on AGNs with asymmetric H$\beta$ emission \citep[][]{Zastrocky2024}, we argue that the BLR of PG~1302-102 shows deviations from a purely spherical or disk-like emissivity distribution. We do not exclude the recent merger of two separate BLRs being a possible cause of such perturbations. However, given the absence of variability in the BELs emission, we argue that self-gravitating instabilities are more likely to be an explanation for the emission coming from the BLR. 

The perturbed BLR scenario is also supported by its capability of simultaneously reproducing the peculiar emission profiles of the H$\gamma$, H$\beta$, and the observed flux excess at $\lambda \simeq 5000\AA$. Indeed, the analysis presented in Sec.~\ref{sec:multi_gauss} of the spectrum of PG~1302-102 revealed an additional component, which, if associated with the H$\beta$, demonstrates high redshift ($\Delta\mu_{\rm{VB}}\simeq4000~\rm{km/s}$) and a large broadening ($\rm{FWHM}_{\rm{VB}}=12600~\rm{km/s}$). This is the first time such component has been detected in PG~1302-102, however, the same feature has already been observed in different quasars \citep[e.g., class B1 and B1+ quasar,][]{Sulentic2002}. One possibility could be this component not being associated with the H$\beta$ but with the emission from the Fe. In our approach, we fixed the ratios of the different Fe multiplets from \cite{Veron_cetty_2004}. Different conditions and geometries of the BLR and of the ionizing source might play a role in changing the multiplet ratios without needing any further components. In the case where the component is associated with the H$\beta$, gravitational redshift from gas close to the SMBH might be a possible explanation. In such case, assuming only gravitational redshift, we estimated a distance of $\simeq 70$ gravitational radii to get a redshift of $\simeq 4000~\rm{km/s}$ (i.e., the peak of the very broad component). It is not clear if, at such small distances from the BH, there are the physical conditions to efficiently produce Balmer emission lines. It is also unclear whether this feature should be expected in all Balmer lines, as our analysis indicates the lack of any very redshifted component in the H$\gamma$. As was already anticipated in Sec.~\ref{sec:disck+spiral}, the disk-like BLR model with the spiral perturbation can capture and explain the asymmetric emission of the H$\beta$ and H$\gamma$ close to the bulk of the line (i.e., $\Delta\mu_B\simeq450~\rm{km/s}$) and the excess (or the lack of it in the case of the H$\gamma$) at much longer wavelengths. This provides not only a self-consistent and physically motivated explanation for the complex nature of the BLR of PG~1302-102 but also a viable explanation for the whole class of quasar showing the same very redshifted component, which, until now, has only been investigated empirically.

Finally, the origin of the BLR disturbance might be connected to outflows. Our analysis has revealed the presence of ionized outflows on NLR scales in PG~1302-102. Few works have reported evidence for a connection between outflows in broad and narrow line regions as the two may be connected and part of the same feedback mechanism \citep[][]{Vietri2020}. Therefore, it might be possible that the redshifted components of the BLR are partially tracing the radial motion of the gas or that the disturbance in the disk originates from feedback from the central AGN. In this case, due to dust absorption and obscuration, it is generally expected to preferentially observe outflows instead of inflows. If AGN winds are really the cause of the observed emission profile, the fact that we detect only a redshifted component without any blueshifted counterpart at least implies quite a peculiar configuration and dynamics of the gas and the dust in the BLR of PG~1302-102. Only precise and long monitoring reverberation mapping campaigns will shed light on the real nature of PG~1302-102 either by characterizing its BLR transfer function or by testing the binary hypothesis against newly proposed models \citep[][Bertassi et al. in prep.]{Dotti2022,Dotti2023}.

\section{Summary and conclusions}
\label{sec:conlcusion}
In this work, we conducted the first fully Bayesian analysis on the high-resolution ESPRESSO spectrum of the binary black hole candidate PG~1302-102. Our Bayesian approach, based on the nested sampling algorithm, measures the posterior probability distribution of all the model parameters and the evidence of the model, which we use to assess if an emission component is statistically required. Our methodology coupled with the high quality of the data (\fr{$\rm{R}\simeq138,000$}) allowed us to confirm (for the H$\beta$) and detect (for the H$\gamma$) the presence of a redshifted broad emission component  ($\Delta\mu\simeq450\rm{km/s}$, $\rm{FWHM_B} = 3981 \rm{km/s}$) for which we provided accurate measurements of its properties. We also detected an additional Gaussian component that, if associated with the H$\beta$, implies high redshift ($\Delta\mu_{\rm{VB}}\simeq4000\rm{km/s}$) and a large broadening ($\rm{FWHM}_{\rm{VB}}=12600\rm{km/s}$); notably the same component is not detected in the H$\gamma$.
Finally, we discovered and characterized the presence of blueshifted broad component ($\Delta\mu \simeq -350\rm{km/s}$, $\rm{FWHM_O}=991\rm{km/s}$) in all the analyzed narrow lines (H$\beta$, H$\gamma$, and [$\rm{OIII}$]) that we explained as a gas outflow.  From the width of the BELs we estimated a dynamical mass of $M_{\rm{BH}}\simeq10^{8.8}~\rm{M_\sun}$ for PG~1302-102.  


We discussed two scenarios as viable explanations for the complex emission of the Balmer lines. A possibility is the presence of a black hole binary carrying around the BLR during the orbital motion. In this case, assuming a period of the binary of $\simeq 5.5\rm{yr}$ \citep[][]{Zhu2020} and the typical luminosity-radius relations with their uncertainties for the average BLR size \citep[][]{bentz09b,Shen2024,Wang2024}, we conclude that the BLRs of the individual black holes should be either already settled in a single BLR or in the process of merging and, therefore highly disturbed and truncated. The alternative explanation assumed the BLR to be a disk orbiting in Keplerian circular motion around a single SMBH with a non-axisymmetric emissivity distribution modeled as a spiral overdensity. We showed that accounting for a perturbation in the BLR, a disk-BLR model well-fitted the data. This approach simultaneously explains the asymmetric emission of the H$\beta$ and H$\gamma$ close to the bulk of the line and the flux excess (or the lack of it in the case of the H$\gamma$) at much longer
wavelengths. A comparison of the Bayesian evidence of the spiral perturbation model with a simpler multi-Gaussian model showed a clear statistical preference for the former. This strongly suggests that the BLR of PG~1302-102 is a complex system, demonstrating that, when feasible, geometrical and dynamical modeling of the BLR should always be preferred over an empirical superposition of unphysical components. Our approach could indeed be extended to other sources, possibly explaining all those objects showing emission line features similar to that of PG~1302-102.

Still, the physical origins of the perturbation are unclear and a connection with the possible presence of a black hole binary cannot be ruled out. However, given the growing evidence from theoretical \citep[][]{Wang2022} and observational \citep[][]{Zastrocky2024} works demonstrating the common presence of disturbed BLRs, we argue that an origin related to self-gravitating instabilities may be more plausible. Future spectroscopic monitoring of PG~1302-102 is surely needed to better characterize the nature of this source and the structure of its BLR.


\section*{Acknowledgements}
\fr{We thank the anonymous referee for their comments and suggestions that helped us to improve the quality of the paper.}

FR acknowledges the support from the Next Generation EU funds within the National Recovery and Resilience Plan (PNRR), Mission 4 - Education and Research, Component 2 - From Research to Business (M4C2), Investment Line 3.1 - Strengthening and creation of Research Infrastructures, Project IR0000012 – “CTA+ - Cherenkov Telescope Array Plus.

MD acknowledge funding from MIUR under the grant
PRIN 2017-MB8AEZ, financial support from ICSC – Centro Nazionale di Ricerca in High Performance Computing, Big Data and Quantum Computing, funded by European Union – NextGenerationEU, and support by the Italian Ministry for Research and University (MUR) under Grant 'Progetto Dipartimenti di Eccellenza 2023-2027' (BiCoQ).
We acknowledge a financial contribution from the Bando Ricerca Fondamentale INAF 2022 Large Grant, {\textit{Dual and binary supermassive black holes in the multi-messenger era: from galaxy mergers to gravitational waves.}}
This work is based on public data released from the ESPRESSO commissioning observations under Programme ID 60.A-9128
The data underlying this article will be shared on reasonable request to the corresponding author. 





\bibliographystyle{aa} 
\bibliography{main}

\providecommand{\pasp}{Publications of the Astronomical Society of the Pacific} \providecommand{\cjaa}{Chinese Journal of Astronomy~\& Astrophysics} \providecommand{\mnras}{MNRAS} \providecommand{\aap}{Astronomy and Astrophysics} \providecommand{\apj}{ApJ} \providecommand{\apjs}{Astrophysical Journal, Supplement} \providecommand{\araa}{Annual Review of Astron and Astrophys} \providecommand{\apjl}{Astrophysical Journal Letters} \providecommand{\aj}{Astronomical Journal}
\begin{thebibliography}{87}
\expandafter\ifx\csname natexlab\endcsname\relax\def\natexlab#1{#1}\fi

\bibitem[{{Ackermann} {et~al.}(2015){Ackermann}, {Ajello}, {Albert}, {Atwood}, {Baldini}, {Ballet}, {Barbiellini}, {Bastieri}, {Becerra Gonzalez}, {Bellazzini}, {Bissaldi}, {Blandford}, {Bloom}, {Bonino}, {Bottacini}, {Bregeon}, {Bruel}, {Buehler}, {Buson}, {Caliandro}, {Cameron}, {Caputo}, {Caragiulo}, {Caraveo}, {Cavazzuti}, {Cecchi}, {Chekhtman}, {Chiang}, {Chiaro}, {Ciprini}, {Cohen-Tanugi}, {Conrad}, {Cutini}, {D'Ammando}, {de Angelis}, {de Palma}, {Desiante}, {Di Venere}, {Dom{\'{\i}}nguez}, {Drell}, {Favuzzi}, {Fegan}, {Ferrara}, {Focke}, {Fuhrmann}, {Fukazawa}, {Fusco}, {Gargano}, {Gasparrini}, {Giglietto}, {Giommi}, {Giordano}, {Giroletti}, {Godfrey}, {Green}, {Grenier}, {Grove}, {Guiriec}, {Harding}, {Hays}, {Hewitt}, {Hill}, {Horan}, {Jogler}, {J{\'o}hannesson}, {Johnson}, {Kamae}, {Kuss}, {Larsson}, {Latronico}, {Li}, {Li}, {Longo}, {Loparco}, {Lott}, {Lovellette}, {Lubrano}, {Magill}, {Maldera}, {Manfreda}, {Max-Moerbeck}, {Mayer}, {Mazziotta}, {McEnery}, {Michelson}, {Mizuno}, {Monzani},
  {Morselli}, {Moskalenko}, {Murgia}, {Nuss}, {Ohno}, {Ohsugi}, {Ojha}, {Omodei}, {Orlando}, {Ormes}, {Paneque}, {Pearson}, {Perkins}, {Perri}, {Pesce-Rollins}, {Petrosian}, {Piron}, {Pivato}, {Porter}, {Rain{\`o}}, {Rando}, {Razzano}, {Readhead}, {Reimer}, {Reimer}, {Schulz}, {Sgr{\`o}}, {Siskind}, {Spada}, {Spandre}, {Spinelli}, {Suson}, {Takahashi}, {Thayer}, {Thompson}, {Tibaldo}, {Torres}, {Tosti}, {Troja}, {Uchiyama}, {Vianello}, {Wood}, {Wood}, {Zimmer}, {Berdyugin}, {Corbet}, {Hovatta}, {Lindfors}, {Nilsson}, {Reinthal}, {Sillanp{\"a}{\"a}}, {Stamerra}, {Takalo}, \& {Valtonen}}]{Ackermann15}
{Ackermann}, M., {Ajello}, M., {Albert}, A., {et~al.} 2015, \apjl, 813, L41

\bibitem[{{Agazie} {et~al.}(2024){Agazie}, {Antoniadis}, {Anumarlapudi}, {Archibald}, {Arumugam}, {Arumugam}, {Arzoumanian}, {Askew}, {Babak}, {Bagchi}, {Bailes}, {Bak Nielsen}, {Baker}, {Bassa}, {Bathula}, {B{\'e}csy}, {Berthereau}, {Bhat}, {Blecha}, {Bonetti}, {Bortolas}, {Brazier}, {Brook}, {Burgay}, {Burke-Spolaor}, {Burnette}, {Caballero}, {Cameron}, {Case}, {Chalumeau}, {Champion}, {Chanlaridis}, {Charisi}, {Chatterjee}, {Chatziioannou}, {Cheeseboro}, {Chen}, {Chen}, {Cognard}, {Cohen}, {Coles}, {Cordes}, {Cornish}, {Crawford}, {Cromartie}, {Crowter}, {Cury{\l}o}, {Cutler}, {Dai}, {Dandapat}, {Deb}, {DeCesar}, {DeGan}, {Demorest}, {Deng}, {Desai}, {Desvignes}, {Dey}, {Dhanda-Batra}, {Di Marco}, {Dolch}, {Drachler}, {Dwivedi}, {Ellis}, {Falxa}, {Feng}, {Ferdman}, {Ferrara}, {Fiore}, {Fonseca}, {Franchini}, {Freedman}, {Gair}, {Garver-Daniels}, {Gentile}, {Gersbach}, {Glaser}, {Good}, {Goncharov}, {Gopakumar}, {Graikou}, {Griessmeier}, {Guillemot}, {G{\"u}ltekin}, {Guo}, {Gupta}, {Grunthal}, {Hazboun},
  {Hisano}, {Hobbs}, {Hourihane}, {Hu}, {Iraci}, {Islo}, {Izquierdo-Villalba}, {Jang}, {Jawor}, {Janssen}, {Jennings}, {Jessner}, {Johnson}, {Jones}, {Joshi}, {Kaiser}, {Kaplan}, {Kapur}, {Kareem}, {Karuppusamy}, {Keane}, {Keith}, {Kelley}, {Kerr}, {Key}, {Kharbanda}, {Kikunaga}, {Klein}, {Kolhe}, {Kramer}, {Krishnakumar}, {Kulkarni}, {Laal}, {Lackeos}, {Lam}, {Lamb}, {Larsen}, {Lazio}, {Lee}, {Levin}, {Lewandowska}, {Littenberg}, {Liu}, {Liu}, {Liu}, {Lommen}, {Lorimer}, {Lower}, {Luo}, {Luo}, {Lynch}, {Lyne}, {Ma}, {Maan}, {Madison}, {Main}, {Manchester}, {Mandow}, {Mattson}, {McEwen}, {McKee}, {McLaughlin}, {McMann}, {Meyers}, {Meyers}, {Mickaliger}, {Miles}, {Mingarelli}, {Mitridate}, {Natarajan}, {Nathan}, {Ng}, {Nice}, {Ni{\c{t}}u}, {Nobleson}, {Ocker}, {Olum}, {Os{\l}owski}, {Paladi}, {Parthasarathy}, {Pennucci}, {Perera}, {Perrodin}, {Petiteau}, {Petrov}, {Pol}, {Porayko}, {Possenti}, {Prabu}, {Quelquejay Leclere}, {Radovan}, {Rana}, {Ransom}, {Ray}, {Reardon}, {Rogers}, {Romano}, {Russell},
  {Samajdar}, {Sanidas}, {Sardesai}, {Schmiedekamp}, {Schmiedekamp}, {Schmitz}, {Schult}, {Sesana}, {Shaifullah}, {Shannon}, {Shapiro-Albert}, {Siemens}, {Simon}, {Singha}, {Siwek}, {Speri}, {Spiewak}, {Srivastava}, {Stairs}, {Stappers}, {Stinebring}, {Stovall}, {Sun}, {Surnis}, {Susarla}, {Susobhanan}, {Swiggum}, {Takahashi}, {Tarafdar}, {Taylor}, {Taylor}, {Theureau}, {Thrane}, {Thyagarajan}, {Tiburzi}, {Toomey}, {Turner}, {Unal}, {Vallisneri}, {van der Wateren}, {van Haasteren}, {Vecchio}, {Venkatraman Krishnan}, {Verbiest}, {Vigeland}, {Wahl}, {Wang}, {Wang}, {Witt}, {Wang}, {Wang}, {Wayt}, {Wu}, {Young}, {Zhang}, {Zhang}, {Zhu}, {Zic}, \& {International Pulsar Timing Array Collaboration}}]{Agazie23}
{Agazie}, G., {Antoniadis}, J., {Anumarlapudi}, A., {et~al.} 2024, \apj, 966, 105

\bibitem[{{Amaro-Seoane} {et~al.}(2023){Amaro-Seoane}, {Andrews}, {Arca Sedda}, {Askar}, {Baghi}, {Balasov}, {Bartos}, {Bavera}, {Bellovary}, {Berry}, {Berti}, {Bianchi}, {Blecha}, {Blondin}, {Bogdanovi{\'c}}, {Boissier}, {Bonetti}, {Bonoli}, {Bortolas}, {Breivik}, {Capelo}, {Caramete}, {Cattorini}, {Charisi}, {Chaty}, {Chen}, {Chru{\'s}li{\'n}ska}, {Chua}, {Church}, {Colpi}, {D'Orazio}, {Danielski}, {Davies}, {Dayal}, {De Rosa}, {Derdzinski}, {Destounis}, {Dotti}, {Dutan}, {Dvorkin}, {Fabj}, {Foglizzo}, {Ford}, {Fouvry}, {Franchini}, {Fragos}, {Fryer}, {Gaspari}, {Gerosa}, {Graziani}, {Groot}, {Habouzit}, {Haggard}, {Haiman}, {Han}, {Istrate}, {Johansson}, {Khan}, {Kimpson}, {Kokkotas}, {Kong}, {Korol}, {Kremer}, {Kupfer}, {Lamberts}, {Larson}, {Lau}, {Liu}, {Lloyd-Ronning}, {Lodato}, {Lupi}, {Ma}, {Maccarone}, {Mandel}, {Mangiagli}, {Mapelli}, {Mathis}, {Mayer}, {McGee}, {McKernan}, {Miller}, {Mota}, {Mumpower}, {Nasim}, {Nelemans}, {Noble}, {Pacucci}, {Panessa}, {Paschalidis}, {Pfister}, {Porquet},
  {Quenby}, {Ricarte}, {R{\"o}pke}, {Regan}, {Rosswog}, {Ruiter}, {Ruiz}, {Runnoe}, {Schneider}, {Schnittman}, {Secunda}, {Sesana}, {Seto}, {Shao}, {Shapiro}, {Sopuerta}, {Stone}, {Suvorov}, {Tamanini}, {Tamfal}, {Tauris}, {Temmink}, {Tomsick}, {Toonen}, {Torres-Orjuela}, {Toscani}, {Tsokaros}, {Unal}, {V{\'a}zquez-Aceves}, {Valiante}, {van Putten}, {van Roestel}, {Vignali}, {Volonteri}, {Wu}, {Younsi}, {Yu}, {Zane}, {Zwick}, {Antonini}, {Baibhav}, {Barausse}, {Bonilla Rivera}, {Branchesi}, {Branduardi-Raymont}, {Burdge}, {Chakraborty}, {Cuadra}, {Dage}, {Davis}, {de Mink}, {Decarli}, {Doneva}, {Escoffier}, {Gandhi}, {Haardt}, {Lousto}, {Nissanke}, {Nordhaus}, {O'Shaughnessy}, {Portegies Zwart}, {Pound}, {Schussler}, {Sergijenko}, {Spallicci}, {Vernieri}, \& {Vigna-G{\'o}mez}}]{lisa2}
{Amaro-Seoane}, P., {Andrews}, J., {Arca Sedda}, M., {et~al.} 2023, Living Reviews in Relativity, 26, 2

\bibitem[{{Amaro-Seoane} {et~al.}(2017){Amaro-Seoane}, {Audley}, {Babak}, {Baker}, {Barausse}, {Bender}, {Berti}, {Binetruy}, {Born}, {Bortoluzzi}, {Camp}, {Caprini}, {Cardoso}, {Colpi}, {Conklin}, {Cornish}, {Cutler}, {Danzmann}, {Dolesi}, {Ferraioli}, {Ferroni}, {Fitzsimons}, {Gair}, {Gesa Bote}, {Giardini}, {Gibert}, {Grimani}, {Halloin}, {Heinzel}, {Hertog}, {Hewitson}, {Holley-Bockelmann}, {Hollington}, {Hueller}, {Inchauspe}, {Jetzer}, {Karnesis}, {Killow}, {Klein}, {Klipstein}, {Korsakova}, {Larson}, {Livas}, {Lloro}, {Man}, {Mance}, {Martino}, {Mateos}, {McKenzie}, {McWilliams}, {Miller}, {Mueller}, {Nardini}, {Nelemans}, {Nofrarias}, {Petiteau}, {Pivato}, {Plagnol}, {Porter}, {Reiche}, {Robertson}, {Robertson}, {Rossi}, {Russano}, {Schutz}, {Sesana}, {Shoemaker}, {Slutsky}, {Sopuerta}, {Sumner}, {Tamanini}, {Thorpe}, {Troebs}, {Vallisneri}, {Vecchio}, {Vetrugno}, {Vitale}, {Volonteri}, {Wanner}, {Ward}, {Wass}, {Weber}, {Ziemer}, \& {Zweifel}}]{lisa1}
{Amaro-Seoane}, P., {Audley}, H., {Babak}, S., {et~al.} 2017, arXiv e-prints, arXiv:1702.00786

\bibitem[{{Begelman} {et~al.}(1980){Begelman}, {Blandford}, \& {Rees}}]{BBR80}
{Begelman}, M.~C., {Blandford}, R.~D., \& {Rees}, M.~J. 1980, \nat, 287, 307

\bibitem[{{Bentz} {et~al.}(2010){Bentz}, {Horne}, {Barth}, {Bennert}, {Canalizo}, {Filippenko}, {Gates}, {Malkan}, {Minezaki}, {Treu}, {Woo}, \& {Walsh}}]{Bentz2010}
{Bentz}, M.~C., {Horne}, K., {Barth}, A.~J., {et~al.} 2010, \apjl, 720, L46

\bibitem[{{Bentz} {et~al.}(2009){Bentz}, {Peterson}, {Netzer}, {Pogge}, \& {Vestergaard}}]{bentz09b}
{Bentz}, M.~C., {Peterson}, B.~M., {Netzer}, H., {Pogge}, R.~W., \& {Vestergaard}, M. 2009, \apj, 697, 160

\bibitem[{{Boroson} \& {Green}(1992)}]{Boroson1992}
{Boroson}, T.~A. \& {Green}, R.~F. 1992, \apjs, 80, 109

\bibitem[{{Britzen} {et~al.}(2018){Britzen}, {Fendt}, {Witzel}, {Qian}, {Pashchenko}, {Kurtanidze}, {Zajacek}, {Martinez}, {Karas}, {Aller}, {Aller}, {Eckart}, {Nilsson}, {Ar{\'e}valo}, {Cuadra}, {Subroweit}, \& {Witzel}}]{Britzen2018}
{Britzen}, S., {Fendt}, C., {Witzel}, G., {et~al.} 2018, \mnras, 478, 3199

\bibitem[{{Burke-Spolaor}(2011)}]{BurkeSpolaor11}
{Burke-Spolaor}, S. 2011, \mnras, 410, 2113

\bibitem[{{Calderone} {et~al.}(2017){Calderone}, {Nicastro}, {Ghisellini}, {Dotti}, {Sbarrato}, {Shankar}, \& {Colpi}}]{Calderone2017}
{Calderone}, G., {Nicastro}, L., {Ghisellini}, G., {et~al.} 2017, \mnras, 472, 4051

\bibitem[{{Charisi} {et~al.}(2016){Charisi}, {Bartos}, {Haiman}, {Price-Whelan}, {Graham}, {Bellm}, {Laher}, \& {M{\'a}rka}}]{Charisi16}
{Charisi}, M., {Bartos}, I., {Haiman}, Z., {et~al.} 2016, \mnras, 463, 2145

\bibitem[{{Chen} {et~al.}(2020){Chen}, {Liu}, {Liao}, {Holgado}, {Guo}, {Gruendl}, {Morganson}, {Shen}, {Zhang}, {Abbott}, {Aguena}, {Allam}, {Avila}, {Bertin}, {Bhargava}, {Brooks}, {Burke}, {Carnero Rosell}, {Carollo}, {Carrasco Kind}, {Carretero}, {Costanzi}, {da Costa}, {Davis}, {De Vicente}, {Desai}, {Diehl}, {Doel}, {Everett}, {Flaugher}, {Friedel}, {Frieman}, {Garc{\'\i}a-Bellido}, {Gaztanaga}, {Glazebrook}, {Gruen}, {Gutierrez}, {Hinton}, {Hollowood}, {James}, {Kim}, {Kuehn}, {Kuropatkin}, {Lewis}, {Lidman}, {Lima}, {Maia}, {March}, {Marshall}, {Menanteau}, {Miquel}, {Palmese}, {Paz-Chinch{\'o}n}, {Plazas}, {Sanchez}, {Schubnell}, {Serrano}, {Sevilla-Noarbe}, {Smith}, {Suchyta}, {Swanson}, {Tarle}, {Tucker}, {Norbert Varga}, \& {Walker}}]{Chen+2020}
{Chen}, Y.-C., {Liu}, X., {Liao}, W.-T., {et~al.} 2020, \mnras, 499, 2245

\bibitem[{{Covino} {et~al.}(2020){Covino}, {Landoni}, {Sandrinelli}, \& {Treves}}]{Covinoetal2020}
{Covino}, S., {Landoni}, M., {Sandrinelli}, A., \& {Treves}, A. 2020, \apj, 895, 122

\bibitem[{{Covino} {et~al.}(2019){Covino}, {Sandrinelli}, \& {Treves}}]{Covinoetal2019}
{Covino}, S., {Sandrinelli}, A., \& {Treves}, A. 2019, \mnras, 482, 1270

\bibitem[{{Cupani} {et~al.}(2016){Cupani}, {D'Odorico}, {Cristiani}, {Gonz{\'a}lez-Hern{\'a}ndez}, {Lovis}, {Sousa}, {Calderone}, {Cirami}, {Di Marcantonio}, \& {M{\'e}gevand}}]{cupani+2016}
{Cupani}, G., {D'Odorico}, V., {Cristiani}, S., {et~al.} 2016, in Society of Photo-Optical Instrumentation Engineers (SPIE) Conference Series, Vol. 9913, Software and Cyberinfrastructure for Astronomy IV, ed. G.~{Chiozzi} \& J.~C. {Guzman}, 99131T

\bibitem[{{Davelaar} \& {Haiman}(2022{\natexlab{a}})}]{davelaar22b}
{Davelaar}, J. \& {Haiman}, Z. 2022{\natexlab{a}}, \prd, 105, 103010

\bibitem[{{Davelaar} \& {Haiman}(2022{\natexlab{b}})}]{davelaar22a}
{Davelaar}, J. \& {Haiman}, Z. 2022{\natexlab{b}}, \prl, 128, 191101

\bibitem[{{De Rosa} {et~al.}(2019){De Rosa}, {Vignali}, {Bogdanovi{\'c}}, {Capelo}, {Charisi}, {Dotti}, {Husemann}, {Lusso}, {Mayer}, {Paragi}, {Runnoe}, {Sesana}, {Steinborn}, {Bianchi}, {Colpi}, {del Valle}, {Frey}, {Gab{\'a}nyi}, {Giustini}, {Guainazzi}, {Haiman}, {Herrera Ruiz}, {Herrero-Illana}, {Iwasawa}, {Komossa}, {Lena}, {Loiseau}, {Perez-Torres}, {Piconcelli}, \& {Volonteri}}]{Derosa2019}
{De Rosa}, A., {Vignali}, C., {Bogdanovi{\'c}}, T., {et~al.} 2019, \nar, 86, 101525

\bibitem[{{D'Orazio} \& {Di Stefano}(2018)}]{doraziolense18}
{D'Orazio}, D.~J. \& {Di Stefano}, R. 2018, \mnras, 474, 2975

\bibitem[{{D'Orazio} {et~al.}(2015){D'Orazio}, {Haiman}, \& {Schiminovich}}]{DHS15}
{D'Orazio}, D.~J., {Haiman}, Z., \& {Schiminovich}, D. 2015, \nat, 525, 351

\bibitem[{{Dotti} {et~al.}(2022){Dotti}, {Bonetti}, {D'Orazio}, {Haiman}, \& {Ho}}]{Dotti2022}
{Dotti}, M., {Bonetti}, M., {D'Orazio}, D.~J., {Haiman}, Z., \& {Ho}, L.~C. 2022, \mnras, 509, 212

\bibitem[{{Dotti} {et~al.}(2023){Dotti}, {Rigamonti}, {Rinaldi}, {Del Pozzo}, {Decarli}, \& {Buscicchio}}]{Dotti2023}
{Dotti}, M., {Rigamonti}, F., {Rinaldi}, S., {et~al.} 2023, \aap, 680, A69

\bibitem[{{Du} {et~al.}(2018){Du}, {Brotherton}, {Wang}, {Huang}, {Hu}, {Kasper}, {Chick}, {Nguyen}, {Maithil}, {Hand}, {Li}, {Ho}, {Bai}, {Bian}, {Wang}, \& {MAHA Collaboration}}]{Du2018}
{Du}, P., {Brotherton}, M.~S., {Wang}, K., {et~al.} 2018, \apj, 869, 142

\bibitem[{{Eracleous} {et~al.}(2012){Eracleous}, {Boroson}, {Halpern}, \& {Liu}}]{Eracleous12}
{Eracleous}, M., {Boroson}, T.~A., {Halpern}, J.~P., \& {Liu}, J. 2012, \apjs, 201, 23

\bibitem[{{Eracleous} {et~al.}(1997){Eracleous}, {Halpern}, {M. Gilbert}, {Newman}, \& {Filippenko}}]{eracleous97}
{Eracleous}, M., {Halpern}, J.~P., {M. Gilbert}, A., {Newman}, J.~A., \& {Filippenko}, A.~V. 1997, \apj, 490, 216

\bibitem[{{Eracleous} {et~al.}(1995){Eracleous}, {Livio}, {Halpern}, \& {Storchi-Bergmann}}]{Eracleous1995}
{Eracleous}, M., {Livio}, M., {Halpern}, J.~P., \& {Storchi-Bergmann}, T. 1995, \apj, 438, 610

\bibitem[{{Gaskell} \& {Harrington}(2018)}]{Gaskell2018}
{Gaskell}, C.~M. \& {Harrington}, P.~Z. 2018, \mnras, 478, 1660

\bibitem[{{Graham} {et~al.}(2015{\natexlab{a}}){Graham}, {Djorgovski}, {Stern}, {Drake}, {Mahabal}, {Donalek}, {Glikman}, {Larson}, \& {Christensen}}]{Graham15}
{Graham}, M.~J., {Djorgovski}, S.~G., {Stern}, D., {et~al.} 2015{\natexlab{a}}, \mnras, 453, 1562

\bibitem[{{Graham} {et~al.}(2015{\natexlab{b}}){Graham}, {Djorgovski}, {Stern}, {Glikman}, {Drake}, {Mahabal}, {Donalek}, {Larson}, \& {Christensen}}]{Graham2015Nat}
{Graham}, M.~J., {Djorgovski}, S.~G., {Stern}, D., {et~al.} 2015{\natexlab{b}}, \nat, 518, 74

\bibitem[{{Grier} {et~al.}(2017){Grier}, {Pancoast}, {Barth}, {Fausnaugh}, {Brewer}, {Treu}, \& {Peterson}}]{Grier2017}
{Grier}, C.~J., {Pancoast}, A., {Barth}, A.~J., {et~al.} 2017, \apj, 849, 146

\bibitem[{{Grier} {et~al.}(2012){Grier}, {Peterson}, {Pogge}, {Denney}, {Bentz}, {Martini}, {Sergeev}, {Kaspi}, {Minezaki}, {Zu}, {Kochanek}, {Siverd}, {Shappee}, {Stanek}, {Araya Salvo}, {Beatty}, {Bird}, {Bord}, {Borman}, {Che}, {Chen}, {Cohen}, {Dietrich}, {Doroshenko}, {Drake}, {Efimov}, {Free}, {Ginsburg}, {Henderson}, {King}, {Koshida}, {Mogren}, {Molina}, {Mosquera}, {Nazarov}, {Okhmat}, {Pejcha}, {Rafter}, {Shields}, {Skowron}, {Szczygiel}, {Valluri}, \& {van Saders}}]{Grier2012}
{Grier}, C.~J., {Peterson}, B.~M., {Pogge}, R.~W., {et~al.} 2012, \apj, 755, 60

\bibitem[{{Hayasaki} {et~al.}(2008){Hayasaki}, {Mineshige}, \& {Ho}}]{HMH08}
{Hayasaki}, K., {Mineshige}, S., \& {Ho}, L.~C. 2008, \apj, 682, 1134

\bibitem[{{Jackson} {et~al.}(1992){Jackson}, {O'Brien}, {Goad}, {Alloin}, {Axon}, {de Bruyn}, {Clavel}, {Dietrich}, {Gondhalekar}, {van Groningen}, {Kollatschny}, {Laurikainen}, {Lawrence}, {McHardy}, {Penston}, {Perez}, {Perez-Fournon}, {Robinson}, {Stirpe}, {Tadhunter}, {Terlevich}, \& {Wagner}}]{Jackson1992}
{Jackson}, N., {O'Brien}, P.~T., {Goad}, M., {et~al.} 1992, \aap, 262, 17

\bibitem[{Jeffreys(1939)}]{Jeffreys1939}
Jeffreys, H. 1939, Theory of Probability (Oxford, England: Clarendon Press)

\bibitem[{{Ji} {et~al.}(2021){Ji}, {Lu}, {Ge}, {Yan}, \& {Song}}]{Ji2021}
{Ji}, X., {Lu}, Y., {Ge}, J., {Yan}, C., \& {Song}, Z. 2021, \apj, 910, 101

\bibitem[{{Jovanovi{\'c}} {et~al.}(2010){Jovanovi{\'c}}, {Popovi{\'c}}, {Stalevski}, \& {Shapovalova}}]{Jovanovic_2010}
{Jovanovi{\'c}}, P., {Popovi{\'c}}, L.~{\v{C}}., {Stalevski}, M., \& {Shapovalova}, A.~I. 2010, \apj, 718, 168

\bibitem[{{Ju} {et~al.}(2013){Ju}, {Greene}, {Rafikov}, {Bickerton}, \& {Badenes}}]{Ju13}
{Ju}, W., {Greene}, J.~E., {Rafikov}, R.~R., {Bickerton}, S.~J., \& {Badenes}, C. 2013, \apj, 777, 44

\bibitem[{{Jun} {et~al.}(2015){Jun}, {Stern}, {Graham}, {Djorgovski}, {Mainzer}, {Cutri}, {Drake}, \& {Mahabal}}]{Jun2015}
{Jun}, H.~D., {Stern}, D., {Graham}, M.~J., {et~al.} 2015, \apjl, 814, L12

\bibitem[{{Kelly} {et~al.}(2009){Kelly}, {Bechtold}, \& {Siemiginowska}}]{Kelly2009}
{Kelly}, B.~C., {Bechtold}, J., \& {Siemiginowska}, A. 2009, \apj, 698, 895

\bibitem[{{Kharb} {et~al.}(2017){Kharb}, {Lal}, \& {Merritt}}]{Kharb2017}
{Kharb}, P., {Lal}, D.~V., \& {Merritt}, D. 2017, Nature Astronomy, 1, 727

\bibitem[{{Kuhn} {et~al.}(2024){Kuhn}, {Shangguan}, {Davies}, {Man}, {Cao}, {Dexter}, {Eisenhauer}, {F{\"o}rster Schreiber}, {Feuchtgruber}, {Genzel}, {Gillessen}, {H{\"o}nig}, {Lutz}, {Netzer}, {Ott}, {Rabien}, {Santos}, {Shimizu}, {Sturm}, \& {Tacconi}}]{Kuhn2024}
{Kuhn}, L., {Shangguan}, J., {Davies}, R., {et~al.} 2024, \aap, 684, A52

\bibitem[{{Li} {et~al.}(2016){Li}, {Wang}, {Ho}, {Lu}, {Qiu}, {Du}, {Hu}, {Huang}, {Zhang}, {Wang}, \& {Bai}}]{Li2016}
{Li}, Y.-R., {Wang}, J.-M., {Ho}, L.~C., {et~al.} 2016, \apj, 822, 4

\bibitem[{{Li} {et~al.}(2019){Li}, {Wang}, {Zhang}, {Wang}, {Huang}, {Lu}, {Hu}, {Du}, {Bon}, {Ho}, {Bai}, {Bian}, {Yuan}, {Winkler}, {Denissyuk}, {Valiullin}, {Bon}, \& {Popovi{\'c}}}]{Li+2019}
{Li}, Y.-R., {Wang}, J.-M., {Zhang}, Z.-X., {et~al.} 2019, \apjs, 241, 33

\bibitem[{{Liu} {et~al.}(2024){Liu}, {Edelson}, {Hern{\'a}ndez Santisteban}, {Kara}, {Montano}, {Gelbord}, {Horne}, {Barth}, {Cackett}, \& {Kaplan}}]{Liu2024}
{Liu}, T., {Edelson}, R., {Hern{\'a}ndez Santisteban}, J.~V., {et~al.} 2024, \apj, 964, 167

\bibitem[{{Liu} {et~al.}(2019){Liu}, {Gezari}, {Ayers}, {Burgett}, {Chambers}, {Hodapp}, {Huber}, {Kudritzki}, {Metcalfe}, {Tonry}, {Wainscoat}, \& {Waters}}]{LiuGez+2019}
{Liu}, T., {Gezari}, S., {Ayers}, M., {et~al.} 2019, \apj, 884, 36

\bibitem[{{Liu} {et~al.}(2018){Liu}, {Gezari}, \& {Miller}}]{Liu2018Asasn}
{Liu}, T., {Gezari}, S., \& {Miller}, M.~C. 2018, \apjl, 859, L12

\bibitem[{{Marziani} {et~al.}(2009){Marziani}, {Sulentic}, {Stirpe}, {Zamfir}, \& {Calvani}}]{Marziani2009}
{Marziani}, P., {Sulentic}, J.~W., {Stirpe}, G.~M., {Zamfir}, S., \& {Calvani}, M. 2009, \aap, 495, 83

\bibitem[{{Montuori} {et~al.}(2011){Montuori}, {Dotti}, {Colpi}, {Decarli}, \& {Haardt}}]{Montuori11}
{Montuori}, C., {Dotti}, M., {Colpi}, M., {Decarli}, R., \& {Haardt}, F. 2011, \mnras, 412, 26

\bibitem[{{Osterbrock} \& {Ferland}(2006)}]{Osterbrock2006}
{Osterbrock}, D.~E. \& {Ferland}, G.~J. 2006, {Astrophysics of gaseous nebulae and active galactic nuclei}

\bibitem[{{Pancoast} {et~al.}(2014){Pancoast}, {Brewer}, {Treu}, {Park}, {Barth}, {Bentz}, \& {Woo}}]{Pancoast2014}
{Pancoast}, A., {Brewer}, B.~J., {Treu}, T., {et~al.} 2014, \mnras, 445, 3073

\bibitem[{{Pepe} {et~al.}(2021){Pepe}, {Cristiani}, {Rebolo}, {Santos}, {Dekker}, {Cabral}, {Di Marcantonio}, {Figueira}, {Lo Curto}, {Lovis}, {Mayor}, {M{\'e}gevand}, {Molaro}, {Riva}, {Zapatero Osorio}, {Amate}, {Manescau}, {Pasquini}, {Zerbi}, {Adibekyan}, {Abreu}, {Affolter}, {Alibert}, {Aliverti}, {Allart}, {Allende Prieto}, {{\'A}lvarez}, {Alves}, {Avila}, {Baldini}, {Bandy}, {Barros}, {Benz}, {Bianco}, {Borsa}, {Bourrier}, {Bouchy}, {Broeg}, {Calderone}, {Cirami}, {Coelho}, {Conconi}, {Coretti}, {Cumani}, {Cupani}, {D'Odorico}, {Damasso}, {Deiries}, {Delabre}, {Demangeon}, {Dumusque}, {Ehrenreich}, {Faria}, {Fragoso}, {Genolet}, {Genoni}, {G{\'e}nova Santos}, {Gonz{\'a}lez Hern{\'a}ndez}, {Hughes}, {Iwert}, {Kerber}, {Knudstrup}, {Landoni}, {Lavie}, {Lillo-Box}, {Lizon}, {Maire}, {Martins}, {Mehner}, {Micela}, {Modigliani}, {Monteiro}, {Monteiro}, {Moschetti}, {Murphy}, {Nunes}, {Oggioni}, {Oliveira}, {Oshagh}, {Pall{\'e}}, {Pariani}, {Poretti}, {Rasilla}, {Rebord{\~a}o}, {Redaelli}, {Santana Tschudi},
  {Santin}, {Santos}, {S{\'e}gransan}, {Schmidt}, {Segovia}, {Sosnowska}, {Sozzetti}, {Sousa}, {Span{\`o}}, {Su{\'a}rez Mascare{\~n}o}, {Tabernero}, {Tenegi}, {Udry}, \& {Zanutta}}]{Pepe2021}
{Pepe}, F., {Cristiani}, S., {Rebolo}, R., {et~al.} 2021, \aap, 645, A96

\bibitem[{{Popovi{\'c}}(2012)}]{popovic2012}
{Popovi{\'c}}, L.~{\v{C}}. 2012, \nar, 56, 74

\bibitem[{{Raimundo} {et~al.}(2020){Raimundo}, {Vestergaard}, {Goad}, {Grier}, {Williams}, {Peterson}, \& {Treu}}]{Raimundo2020}
{Raimundo}, S.~I., {Vestergaard}, M., {Goad}, M.~R., {et~al.} 2020, \mnras, 493, 1227

\bibitem[{{Rigamonti} {et~al.}(2023){Rigamonti}, {Dotti}, {Covino}, {Haardt}, {Cortese}, {Landoni}, \& {Varisco}}]{Rigamonti2023}
{Rigamonti}, F., {Dotti}, M., {Covino}, S., {et~al.} 2023, \mnras, 525, 1008

\bibitem[{{Rigamonti} {et~al.}(2022){Rigamonti}, {Dotti}, {Covino}, {Haardt}, {Landoni}, {Del Pozzo}, {Lupi}, \& {Zibetti}}]{Rigamonti2022}
{Rigamonti}, F., {Dotti}, M., {Covino}, S., {et~al.} 2022, \mnras, 513, 6111

\bibitem[{{Rodriguez} {et~al.}(2009){Rodriguez}, {Taylor}, {Zavala}, {Pihlstr{\"o}m}, \& {Peck}}]{Rodriguez09}
{Rodriguez}, C., {Taylor}, G.~B., {Zavala}, R.~T., {Pihlstr{\"o}m}, Y.~M., \& {Peck}, A.~B. 2009, \apj, 697, 37

\bibitem[{{Runnoe} {et~al.}(2015){Runnoe}, {Eracleous}, {Mathes}, {Pennell}, {Boroson}, {Sigur{\dh}sson}, {Bogdanovi{\'c}}, {Halpern}, \& {Liu}}]{runnoe15size}
{Runnoe}, J.~C., {Eracleous}, M., {Mathes}, G., {et~al.} 2015, \apjs, 221, 7

\bibitem[{{Saade} {et~al.}(2024){Saade}, {Brightman}, {Stern}, {Connor}, {Djorgovski}, {D'Orazio}, {Ford}, {Graham}, {Haiman}, {Jun}, {Kammoun}, {Kraft}, {McKernan}, {Vikhlinin}, \& {Walton}}]{Saade2024}
{Saade}, M.~L., {Brightman}, M., {Stern}, D., {et~al.} 2024, \apj, 966, 104

\bibitem[{{Sandrinelli} {et~al.}(2016){Sandrinelli}, {Covino}, {Dotti}, \& {Treves}}]{Sandrinelli16}
{Sandrinelli}, A., {Covino}, S., {Dotti}, M., \& {Treves}, A. 2016, \aj, 151, 54

\bibitem[{{Sandrinelli} {et~al.}(2018){Sandrinelli}, {Covino}, {Treves}, {Holgado}, {Sesana}, {Lindfors}, \& {Ramazani}}]{Sandrinelli18}
{Sandrinelli}, A., {Covino}, S., {Treves}, A., {et~al.} 2018, \aap, 615, A118

\bibitem[{{Serafinelli} {et~al.}(2020){Serafinelli}, {Severgnini}, {Braito}, {Della Ceca}, {Vignali}, {Ambrosino}, {Cicone}, {Zaino}, {Dotti}, {Sesana}, {Gianolli}, {Ballo}, {La Parola}, \& {Matzeu}}]{Serafinelli+20}
{Serafinelli}, R., {Severgnini}, P., {Braito}, V., {et~al.} 2020, \apj, 902, 10

\bibitem[{{Severgnini} {et~al.}(2018){Severgnini}, {Cicone}, {Della Ceca}, {Braito}, {Caccianiga}, {Ballo}, {Campana}, {Moretti}, {La Parola}, {Vignali}, {Zaino}, {Matzeu}, \& {Landoni}}]{Severgnini18}
{Severgnini}, P., {Cicone}, C., {Della Ceca}, R., {et~al.} 2018, \mnras, 479, 3804

\bibitem[{{Shen} {et~al.}(2024){Shen}, {Grier}, {Horne}, {Stone}, {Li}, {Yang}, {Homayouni}, {Trump}, {Anderson}, {Brandt}, {Hall}, {Ho}, {Jiang}, {Petitjean}, {Schneider}, {Tao}, {Donnan}, {AlSayyad}, {Bershady}, {Blanton}, {Bizyaev}, {Bundy}, {Chen}, {Davis}, {Dawson}, {Fan}, {Greene}, {Gr{\"o}ller}, {Guo}, {Ibarra-Medel}, {Jiang}, {Keenan}, {Kollmeier}, {Lejoly}, {Li}, {de la Macorra}, {Moe}, {Nie}, {Rossi}, {Smith}, {Tee}, {Weijmans}, {Xu}, {Yue}, {Zhou}, {Zhou}, \& {Zou}}]{Shen2024}
{Shen}, Y., {Grier}, C.~J., {Horne}, K., {et~al.} 2024, \apjs, 272, 26

\bibitem[{{Shen} {et~al.}(2013){Shen}, {Liu}, {Loeb}, \& {Tremaine}}]{Shen13}
{Shen}, Y., {Liu}, X., {Loeb}, A., \& {Tremaine}, S. 2013, \apj, 775, 49

\bibitem[{{Skilling}(2004)}]{Skilling2004}
{Skilling}, J. 2004, in American Institute of Physics Conference Series, Vol. 735, American Institute of Physics Conference Series, ed. R.~{Fischer}, R.~{Preuss}, \& U.~V. {Toussaint}, 395--405

\bibitem[{{Song} {et~al.}(2021){Song}, {Ge}, {Lu}, {Yan}, \& {Ji}}]{Song2021}
{Song}, Z., {Ge}, J., {Lu}, Y., {Yan}, C., \& {Ji}, X. 2021, \aap, 645, A15

\bibitem[{{Storchi-Bergmann} {et~al.}(2003{\natexlab{a}}){Storchi-Bergmann}, {Nemmen da Silva}, {Eracleous}, {Halpern}, {Wilson}, {Filippenko}, {Ruiz}, {Smith}, \& {Nagar}}]{Storchi-Bergmann}
{Storchi-Bergmann}, T., {Nemmen da Silva}, R., {Eracleous}, M., {et~al.} 2003{\natexlab{a}}, \apj, 598, 956

\bibitem[{{Storchi-Bergmann} {et~al.}(2003{\natexlab{b}}){Storchi-Bergmann}, {Nemmen da Silva}, {Eracleous}, {Halpern}, {Wilson}, {Filippenko}, {Ruiz}, {Smith}, \& {Nagar}}]{storchi03}
{Storchi-Bergmann}, T., {Nemmen da Silva}, R., {Eracleous}, M., {et~al.} 2003{\natexlab{b}}, \apj, 598, 956

\bibitem[{{Storchi-Bergmann} {et~al.}(2017){Storchi-Bergmann}, {Schimoia}, {Peterson}, {Elvis}, {Denney}, {Eracleous}, \& {Nemmen}}]{storchi17}
{Storchi-Bergmann}, T., {Schimoia}, J.~S., {Peterson}, B.~M., {et~al.} 2017, \apj, 835, 236

\bibitem[{{Sulentic} {et~al.}(2002){Sulentic}, {Marziani}, {Zamanov}, {Bachev}, {Calvani}, \& {Dultzin-Hacyan}}]{Sulentic2002}
{Sulentic}, J.~W., {Marziani}, P., {Zamanov}, R., {et~al.} 2002, \apjl, 566, L71

\bibitem[{{Tiede} {et~al.}(2024){Tiede}, {Zrake}, {MacFadyen}, \& {Haiman}}]{Tiede2024}
{Tiede}, C., {Zrake}, J., {MacFadyen}, A., \& {Haiman}, Z. 2024, arXiv e-prints, arXiv:2410.03830

\bibitem[{{Tsalmantza} {et~al.}(2011){Tsalmantza}, {Decarli}, {Dotti}, \& {Hogg}}]{Tsalmantza11}
{Tsalmantza}, P., {Decarli}, R., {Dotti}, M., \& {Hogg}, D.~W. 2011, \apj, 738, 20

\bibitem[{{Tsuzuki} {et~al.}(2006){Tsuzuki}, {Kawara}, {Yoshii}, {Oyabu}, {Tanab{\'e}}, \& {Matsuoka}}]{Tsuzuki2006}
{Tsuzuki}, Y., {Kawara}, K., {Yoshii}, Y., {et~al.} 2006, \apj, 650, 57

\bibitem[{{Valtonen} {et~al.}(2008){Valtonen}, {Lehto}, {Nilsson}, {Heidt}, {Takalo}, {Sillanp{\"a}{\"a}}, {Villforth}, {Kidger}, {Poyner}, {Pursimo}, {Zola}, {Wu}, {Zhou}, {Sadakane}, {Drozdz}, {Koziel}, {Marchev}, {Ogloza}, {Porowski}, {Siwak}, {Stachowski}, {Winiarski}, {Hentunen}, {Nissinen}, {Liakos}, \& {Dogru}}]{Valtonen08}
{Valtonen}, M.~J., {Lehto}, H.~J., {Nilsson}, K., {et~al.} 2008, \nat, 452, 851

\bibitem[{{Vaughan} {et~al.}(2016){Vaughan}, {Uttley}, {Markowitz}, {Huppenkothen}, {Middleton}, {Alston}, {Scargle}, \& {Farr}}]{Vaughan2016}
{Vaughan}, S., {Uttley}, P., {Markowitz}, A.~G., {et~al.} 2016, \mnras, 461, 3145

\bibitem[{{Verbiest} {et~al.}(2016){Verbiest}, {Lentati}, {Hobbs}, {van Haasteren}, {Demorest}, {Janssen}, {Wang}, {Desvignes}, {Caballero}, {Keith}, {Champion}, {Arzoumanian}, {Babak}, {Bassa}, {Bhat}, {Brazier}, {Brem}, {Burgay}, {Burke-Spolaor}, {Chamberlin}, {Chatterjee}, {Christy}, {Cognard}, {Cordes}, {Dai}, {Dolch}, {Ellis}, {Ferdman}, {Fonseca}, {Gair}, {Garver-Daniels}, {Gentile}, {Gonzalez}, {Graikou}, {Guillemot}, {Hessels}, {Jones}, {Karuppusamy}, {Kerr}, {Kramer}, {Lam}, {Lasky}, {Lassus}, {Lazarus}, {Lazio}, {Lee}, {Levin}, {Liu}, {Lynch}, {Lyne}, {Mckee}, {McLaughlin}, {McWilliams}, {Madison}, {Manchester}, {Mingarelli}, {Nice}, {Os{\l}owski}, {Palliyaguru}, {Pennucci}, {Perera}, {Perrodin}, {Possenti}, {Petiteau}, {Ransom}, {Reardon}, {Rosado}, {Sanidas}, {Sesana}, {Shaifullah}, {Shannon}, {Siemens}, {Simon}, {Smits}, {Spiewak}, {Stairs}, {Stappers}, {Stinebring}, {Stovall}, {Swiggum}, {Taylor}, {Theureau}, {Tiburzi}, {Toomey}, {Vallisneri}, {van Straten}, {Vecchio}, {Wang}, {Wen}, {You},
  {Zhu}, \& {Zhu}}]{pta}
{Verbiest}, J.~P.~W., {Lentati}, L., {Hobbs}, G., {et~al.} 2016, \mnras, 458, 1267

\bibitem[{{V{\'e}ron-Cetty} {et~al.}(2004){V{\'e}ron-Cetty}, {Joly}, \& {V{\'e}ron}}]{Veron_cetty_2004}
{V{\'e}ron-Cetty}, M.~P., {Joly}, M., \& {V{\'e}ron}, P. 2004, \aap, 417, 515

\bibitem[{{Vietri} {et~al.}(2020){Vietri}, {Mainieri}, {Kakkad}, {Netzer}, {Perna}, {Circosta}, {Harrison}, {Zappacosta}, {Husemann}, {Padovani}, {Bischetti}, {Bongiorno}, {Brusa}, {Carniani}, {Cicone}, {Comastri}, {Cresci}, {Feruglio}, {Fiore}, {Lanzuisi}, {Mannucci}, {Marconi}, {Piconcelli}, {Puglisi}, {Salvato}, {Schramm}, {Schulze}, {Scholtz}, {Vignali}, \& {Zamorani}}]{Vietri2020}
{Vietri}, G., {Mainieri}, V., {Kakkad}, D., {et~al.} 2020, \aap, 644, A175

\bibitem[{{Volonteri} \& {Madau}(2008)}]{Volonteri2008}
{Volonteri}, M. \& {Madau}, P. 2008, \apjl, 687, L57

\bibitem[{{Wang} {et~al.}(2022){Wang}, {Du}, {Songsheng}, \& {Li}}]{Wang2022}
{Wang}, J.-M., {Du}, P., {Songsheng}, Y.-Y., \& {Li}, Y.-R. 2022, \aap, 666, A86

\bibitem[{{Wang} {et~al.}(2017){Wang}, {Greene}, {Ju}, {Rafikov}, {Ruan}, \& {Schneider}}]{Wang17}
{Wang}, L., {Greene}, J.~E., {Ju}, W., {et~al.} 2017, \apj, 834, 129

\bibitem[{{Wang} \& {Woo}(2024)}]{Wang2024}
{Wang}, S. \& {Woo}, J.-H. 2024, \apjs, 275, 13

\bibitem[{{Xin} {et~al.}(2020){Xin}, {Charisi}, {Haiman}, {Schiminovich}, {Graham}, {Stern}, \& {D'Orazio}}]{xin20}
{Xin}, C., {Charisi}, M., {Haiman}, Z., {et~al.} 2020, \mnras, 496, 1683

\bibitem[{{Yu} \& {Lu}(2001)}]{Yu2001}
{Yu}, Q. \& {Lu}, Y. 2001, \aap, 377, 17

\bibitem[{{Zastrocky} {et~al.}(2024){Zastrocky}, {Brotherton}, {Du}, {McLane}, {Olson}, {Dale}, {Kobulnicky}, {Maithil}, {Nguyen}, {Chick}, {Kasper}, {Hand}, {Adelman}, {Carter}, {Murphree}, {Oeur}, {Roth}, {Schonsberg}, {Caradonna}, {Favro}, {Ferguson}, {Gonzalez}, {Hadding}, {Hagler}, {Rogers}, {Stack}, {Chapman}, {Bao}, {Fang}, {Zhai}, {Yang}, {Chen}, {Bai}, {Fu}, {Liu}, {Yao}, {Peng}, {Songsheng}, {Li}, {Bai}, {Hu}, {Xiao}, {Ho}, \& {Wang}}]{Zastrocky2024}
{Zastrocky}, T.~E., {Brotherton}, M.~S., {Du}, P., {et~al.} 2024, \apjs, 272, 29

\bibitem[{{Zhu} \& {Thrane}(2020)}]{Zhu2020}
{Zhu}, X.-J. \& {Thrane}, E. 2020, \apj, 900, 117

\end{thebibliography}



\begin{appendix}

\section{Model with very redshifted and very broad component }
\label{App:model2}

\begin{figure}[h!]
    \centering
    \includegraphics[width=0.5\textwidth,height=0.35\textwidth]{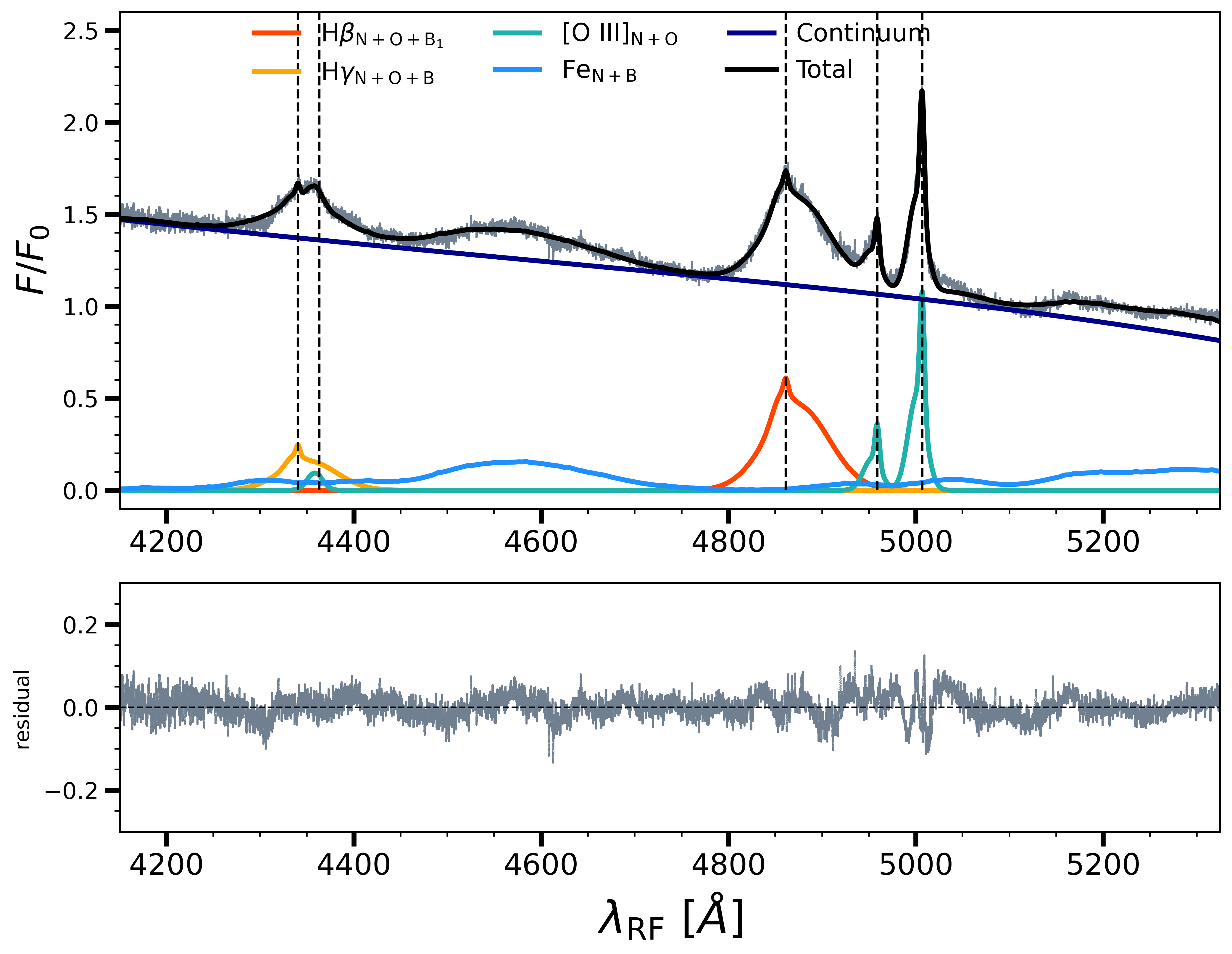}
    \caption{Best-fit result of Model 2 on the ESPRESSO data. The top panel shows the data in gray smoothed with a Gaussian Kernel ($\sigma \simeq 0.07\AA$), the best-fit model (black line), and all the emission components identified by different colors: $\rm{H\gamma}$ in gold, $\rm{H\beta}$ in red, $[\rm{OIII}]$ in cyan, Fe in light-blue and the continuum in dark blue. The vertical dashed lines indicate, the rest-frame emission wavelength of the $\rm{H\gamma}$, $\rm{H\beta}$, and $[\rm{OIII}]$. The bottom panel represents the residuals defined as (data-model)/error, the horizontal dashed black line at the zero level of residuals is plotted to guide the eyes.}
    \label{fig:bf_no_outflow}
\end{figure}

We report in Fig.~\ref{fig:bf_no_outflow} the same plot presented in Fig.~\ref{fig:bf_fiducial} but for a model without the very broad and very redshifted component associated with the H$\beta$ emission line (i.e., Model 2). The strongest differences are observed at $\lambda\simeq5000$. Model 2, without the inclusion of the additional Gaussian component, cannot reproduce the flux excess observed a those wavelengths.

\section{Emissivity distribution for the H$\gamma$ emission line}
\begin{figure}[h!]
    \centering
    \includegraphics[width=0.5\textwidth,height=0.35\textwidth]{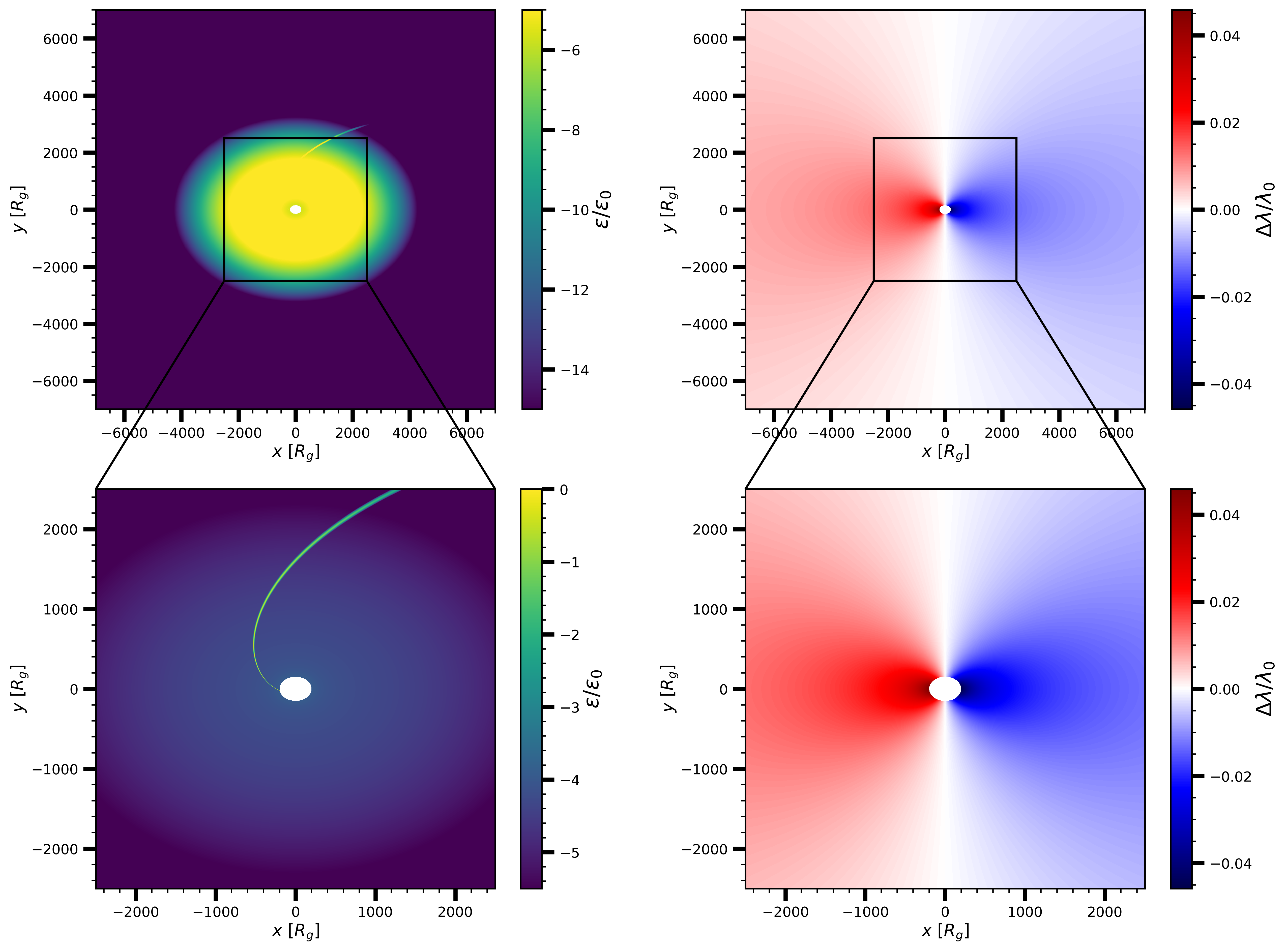}
    \caption{Projected H$\gamma$ emission and Doppler shift of the spiral BLR best-fit. The top left panel is the projected BLR emissivity normalized to the maximum emissivity for the H$\gamma$, while the right top panel represents the Doppler shift of each BLR element. The bottom panels show a zoom of the emissivity (left) and Doppler shift (right) on the central region of the BLR.}
    \label{fig:2D_projected_emissivity_Hgamma}
\end{figure}

We report in Fig.~\ref{fig:2D_projected_emissivity_Hgamma} the projected emissivity for the H$\gamma$ emission line (left) and the Doppler shift (right) at each position of the BLR. The H$\gamma$ emissivity starts to decline at smaller radii compared to that of the H$\beta$ as a consequence of the smaller $\sigma_{\xi_{c,\rm{H\gamma}}}$ that results in a more compact configuration.

\end{appendix}


\label{lastpage}
\end{document}